\documentclass{article}\usepackage[]{graphicx}\usepackage[]{color}
\makeatletter
\def\maxwidth{ %
  \ifdim\Gin@nat@width>\linewidth
    \linewidth
  \else
    \Gin@nat@width
  \fi
}
\makeatother

\definecolor{fgcolor}{rgb}{0.345, 0.345, 0.345}

\usepackage{framed}
\makeatletter
 {\par\unskip\endMakeFramed%
 \at@end@of@kframe}
\makeatother

\definecolor{shadecolor}{rgb}{.97, .97, .97}
\definecolor{messagecolor}{rgb}{0, 0, 0}
\definecolor{warningcolor}{rgb}{1, 0, 1}
\definecolor{errorcolor}{rgb}{1, 0, 0}
\newenvironment{knitrout}{}{} 

\usepackage{alltt}
\usepackage[top=1.0in, bottom=1.0in, left=1.0in, right=1.0in]{geometry}
\usepackage{graphicx}
\usepackage{amsmath}
\usepackage{amssymb}
\usepackage{mathrsfs}
\usepackage{bm}
\usepackage{framed}
\usepackage{lscape}
\usepackage{setspace}
\usepackage{ifxetex}
\usepackage{blkarray}

\IfFileExists{upquote.sty}{\usepackage{upquote}}{}
\begin{document}
\title{Shortfall from Maximum Convexity}
\author{Matthew Ginley\\ Rice University}
\date{}
\maketitle

\begin{abstract}
We review the dynamics of the returns of Leveraged Exchange Traded Funds (LETFs) and propose a new measure of realized volatility: Shortfall from Maximum Convexity.  We show that SMC has a more intuitive interpretation and provides more statistical information compared to the traditionally used sample standard deviation when applied to LETF returns, a dataset where normality and independence do not hold.  \\
\end{abstract}


\clearpage


\section{Dynamics of Leveraged ETF Returns} \label{chapter_etf}

\subsection{Universe of LETFs}

The first Exchange Traded Fund (ETF) began trading in 1993, but by 2015 there are over 1,400 ETFs from over 15 different issuers, with over \$1 trillion invested in ETFs globally \cite{SimpsonInvestopedia}. The existing universe of ETFs covers all major asset classes such as equities, debt, commodities, and real estate, and many ETFs are referenced as the underlying asset of actively traded options.  We will focus on a specific subset of ETFs called Leveraged ETFs, which feature a leverage ratio different than 1:1 as part of their fund structure. \\

The mandate for LETFs is to provide the daily return of an underlying asset (most often an equity index), at the multiple indicated by the leverage ratio.  As of 2015 there are fund offerings featuring leverage ratios of 2:1, -2:1, 3:1, and -3:1.  Going forward, we will simply refer to leverage by its multiple (i.e. 2, -2, 3, -3).  For example, if the underlying index returned 1\% on the day, then LETFs with leverage multiples of 2, -2, 3, and -3 would have a mandate to return 2\%, -2\%, 3\%, and -3\% on the day, respectively.  Positive values indicate long market exposure, while negative multiples indicate short exposure, or betting that the market will decline.  Funds that feature negative leverage multiples are often times referred to as Inverse ETFs, but we will not make this distinction because the returns dynamics are the same except for the obvious negative sign on the leverage term.  \\

In addition to the included leverage, providing investors with the ability to easily acquire short exposure is an important selling point for LETF issuers.  Before these LETFs started trading in 2006 leveraging one's assets or betting against the market required utilizing expensive margin and collateral accounts or navigating the challenges involved with trading options \cite{YatesInvestopedia}.  Investors have since grown to appreciate the conveniences offered by the LETF structure and as of 2015 there are over 100 funds with a total notional value over \$20 billion listed on exchanges worldwide, with Direxion, ProShares, Credit Suisse, and Barclays issuing the majority of them \cite{ETFDatabaseLETFListing}.  For our purposes, we have selected a wide ranging demonstration set of 10 equity indexes referenced as the underlyer by 10 pairs of long and short triple levered ETFs (10 funds with a +3 leverage multiple and 10 with a -3 multiple) issued by Direxion and ProShares.  Our set includes established U.S. market indexes (S\&P 500, Dow Jones Industrial Average, Russell 2000, NASDAQ 100), sector specific indexes (NYSE ARCA Gold Miners, Dow Jones U.S. Financials, S\&P Technology Select), and international equity indexes (MSCI Developed Markets, MSCI Emerging Markets, and Market Vectors Russia). \\

\begin{table}[ht]
\centering
{\scriptsize
\begin{tabular}{ll|ll|ll}
 Issuer & Exposure & LETF & LETF Name & Index & Index Name \\ 
  \hline
\hline
ProShares & Long & FINU & UltraPro Financials & DJUSFN & Dow Jones U.S. Financials Index \\ 
  ProShares & Short & FINZ & UltraPro Short Financials & DJUSFN & Dow Jones U.S. Financials Index \\ 
   \hline
Direxion & Long & NUGT & Daily Gold Miners Bull 3x Shares & GDM & NYSE Arca Gold Miners Index \\ 
  Direxion & Short & DUST & Daily Gold Miners Bear 3x Shares & GDM & NYSE Arca Gold Miners Index \\ 
   \hline
ProShares & Long & UDOW & UltraPro Dow30 & INDU & Dow Jones Industrial Average Index \\ 
  ProShares & Short & SDOW & UltraPro Short Dow30 & INDU & Dow Jones Industrial Average \\ 
   \hline
Direxion & Long & TECL & Daily Technology Bull 3x Shares & IXT & S\&P Technology Select Sector Index \\ 
  Direxion & Short & TECS & Daily Technology Bear 3x Shares & IXT & S\&P Technology Select Sector Index \\ 
   \hline
Direxion & Long & RUSL & Daily Russia Bull 3x Shares & MVRSX & Market Vectors Russia Index \\ 
  Direxion & Short & RUSS & Daily Russia Bear 3x Shares & MVRSX & Market Vectors Russia Index \\ 
   \hline
Direxion & Long & DZK & Daily Developed Markets Bull 3x Shares & MXEA & MSCI EAFE Index \\ 
  Direxion & Short & DPK & Daily Developed Markets Bear 3x Shares & MXEA & MSCI EAFE Index \\ 
   \hline
Direxion & Long & EDC & Daily Emerging Markets Bull 3x Shares & MXEF & MSCI Emerging Markets Index \\ 
  Direxion & Short & EDZ & Daily Emerging Markets Bear 3x Shares & MXEF & MSCI Emerging Markets Index \\ 
   \hline
ProShares & Long & TQQQ & UltraPro QQQ & NDX & NASDAQ-100 Index \\ 
  ProShares & Short & SQQQ & UltraPro Short QQQ & NDX & NASDAQ-100 Index \\ 
   \hline
Direxion & Long & TNA & Daily Small Cap Bull 3x Shares & RTY & Russell 2000 Index \\ 
  Direxion & Short & TZA & Daily Small Cap Bear 3x Shares & RTY & Russell 2000 Index \\ 
   \hline
Direxion & Long & SPXL & Daily S\&P 500 Bull 3x Shares & SPX & S\&P 500 Index \\ 
  Direxion & Short & SPXS & Daily S\&P 500 Bear 3x Shares & SPX & S\&P 500 Index \\ 
  \end{tabular}
}
\caption{Demonstration Set of Selected Pairs of LETFs} 
\label{fig_visindextable}
\end{table}

As their mandate is to provide a multiple of the daily return of the underlying asset, LETFs are intended for short term hedging of market risk.  However, some investors have created buy and hold strategies that incorporate long term positions in LETFs \cite{Fisher2014}.  A primary motivation for our work was what we perceive to be a lack of understanding of the risks involved with holding LETF positions.  As we shall demonstrate later on, the daily compounding of leveraged returns produces much greater volatility than a simple leveraged investment in the index without daily compounding.  \\

\subsection{LETF Returns}

We decompose daily LETF returns with the following time series model ... \\

$R_{LETF,t} = \beta_{LETF} R_{Index,t} - MgmtFee_{LETF} + \epsilon_{LETF,t}$ \\

\begin{tabular}{ll}
$\beta_{LETF}$ & indicated leverage multiple (constant) \\
$MgmtFee_{LETF}$ & indicated rate of daily management expenses (constant) \\
$\epsilon_{LETF,t}$ & tracking error on observation day $t$ (random variable) \\
$R_{Index,t}$ & total return of underlying index on day $t$ (random variable)
\end{tabular}

$ $ \\

The leverage multiple and annualized management expenses are indicated in a given LETF's fund prospectus.  As mentioned, as of 2015 there are commercially available LETFs with leverage multiples of 2, -2, 3, and -3.  All of the management fees indicated on the LETFs in our demonstration set were listed as 0.95\% annually, or a daily rate of approximately 0.375 basis points \cite{DirexionSPXLFactSheet} \cite{DirexionSPXLProspectus} \cite{ProsharesFunds}.  Fees for other LETFs from Direxion and ProShares or other fund issuers are the same or comparable \cite{ETFDatabaseLETFListing}. \\

The daily index return and LETF return usually refer to a total return, or price return in addition to daily dividend yield.  Fortunately, we do not concern ourselves with the calculations involved or the dates of LETF share splits because the Bloomberg Terminal can directly provide a daily returns time series.  All of our work involves only the daily returns series and the leverage and fees constants indicated in fund prospectuses. \\

We include Figures \ref{fig:fig_vishist_error_small} and \ref{fig:fig_vishist_error_large} to illustrate the sampling distributions for the tracking errors ($\epsilon_{LETF,t}$) for our demonstration set of funds.  The 20 funds have been split into a large tracking error group and a small tracking error group for presentation purposes.  The distinction is based on having a min (or max) that is less than -0.025 (or greater than 0.025), or a sample standard deviation greater than 0.01.  While we hope this distinction highlights the scale of possible tracking errors experienced by LETFs, we do not use this distinction anywhere in our modeling.  It is merely used as a rule of thumb for arranging the plots over multiple pages. \\

Overall, it appears as if the LETFs in our set do a reasonable job of hitting their respective performance mandates.  With the exception of LETF Tickers RUSS and FINZ, all of the distributions have a sample mean of -0.0002 or less (-0.02\% or less).  For the small tracking error group, 4 out of 10 have a sample mean of 0.  The LETFs with the longest returns history are SPXL and SPXS, the long and short pair tracking the S\&P 500 index issued by Direxion, for which we have data going back to November 2008.  It is interesting to note that for the 3 pairs of LETFs in the large tracking error group that represent the MVRSX, MXEA, and MXEF indexes, tracking errors in excess of +/- 5\% are not infrequent.  One might suppose that errors of this magnitude would be considered unacceptable by the investing public, but that is not the case.  As of early 2015 all of the LETFs in our demonstration set are still actively traded. \\

\clearpage
\begin{knitrout}
\definecolor{shadecolor}{rgb}{0.969, 0.969, 0.969}\color{fgcolor}\begin{figure}
\includegraphics[width=\maxwidth]{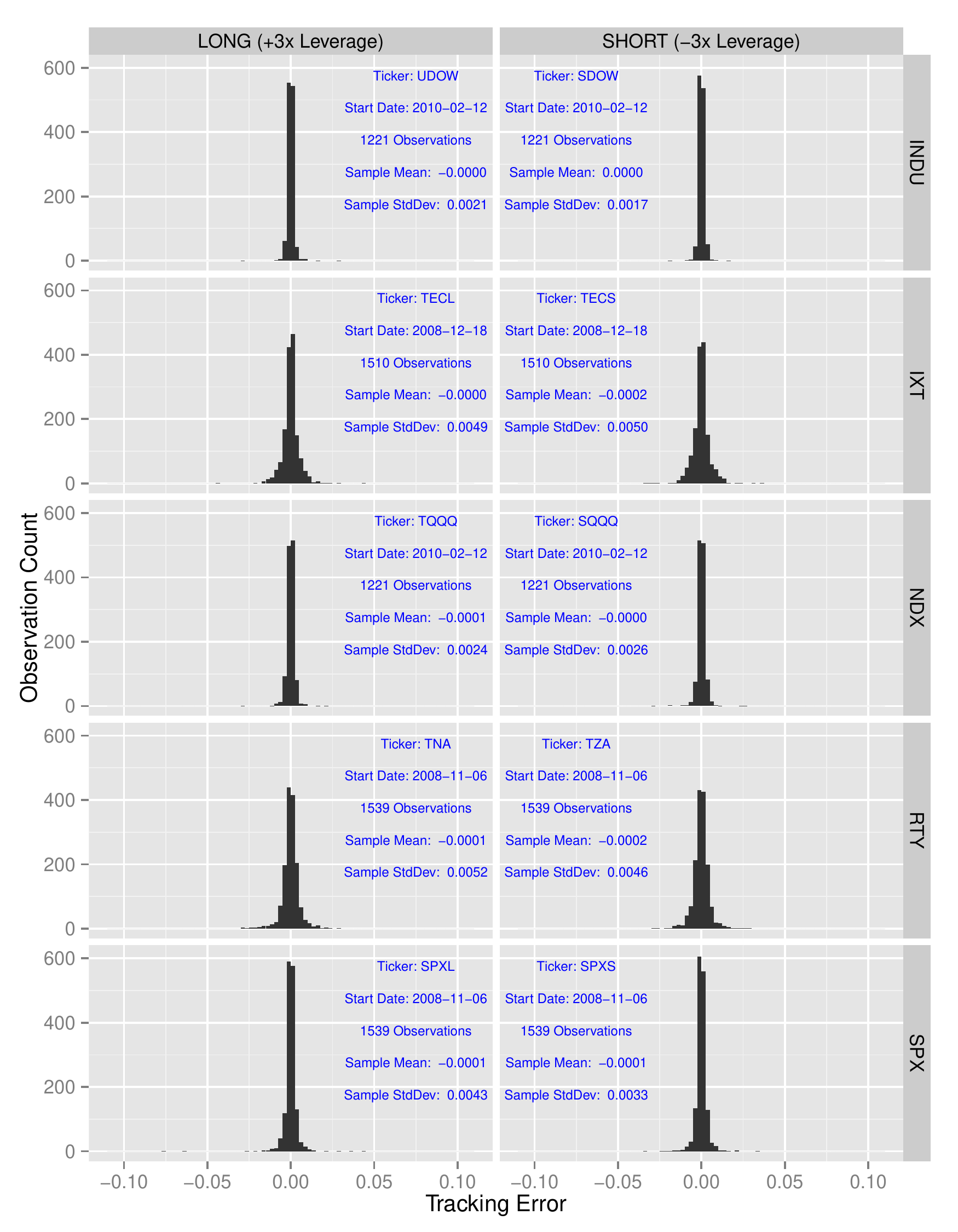} \caption[Histograms of Tracking Error for Small Tracking Error Group]{Histograms of Tracking Error for Small Tracking Error Group}\label{fig:fig_vishist_error_small}
\end{figure}

\end{knitrout}

\clearpage
\begin{knitrout}
\definecolor{shadecolor}{rgb}{0.969, 0.969, 0.969}\color{fgcolor}\begin{figure}
\includegraphics[width=\maxwidth]{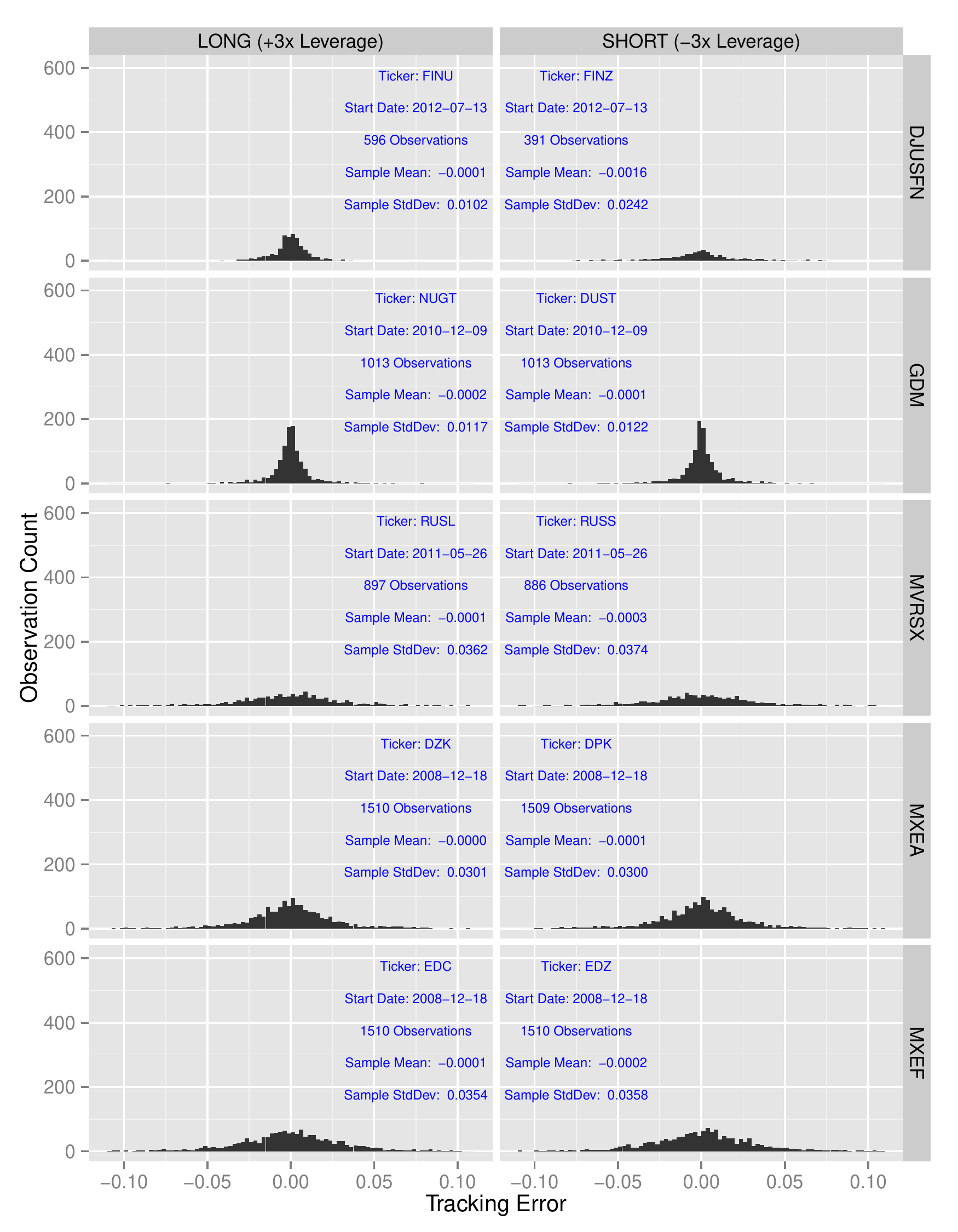} \caption[Histograms of Tracking Error for Large Tracking Error Group]{Histograms of Tracking Error for Large Tracking Error Group}\label{fig:fig_vishist_error_large}
\end{figure}

\end{knitrout}
\clearpage

\subsection{Volatility Drag and Convexity}

%
%
%
%

For the remainder of this section let us ignore specific instruments, fund management fees, and tracking errors, and concern ourselves only with the effects of leveraged compounding of daily returns.  Before proceeding with LETF return dynamics, we need to define the concept of volatility drag.  Traditionally, this refers to the dynamic where the geometric return for the period is less than the arithmetic return \cite{BoucheyNemtchinov2013} ... \\

$\prod_{j = 1}^{p} (1 + R_{j}) - 1 < \sum_{j = 1}^{p} R_j$ \\

This relationship can be easily understood by example.  Say an index experiences a 1\% loss followed by a 1\% gain, then the final compound return (geometric) observed will be negative, while the arithmetic return is clearly 0. \\

$((1 - 0.01) * (1 + 0.01)) - 1 = -0.0001 < 0 = -0.01 + 0.01$ \\

This condition can flip so that the geometric return is greater than the arithmetic return ... \\

$\prod_{j = 1}^{p} (1 + R_{j}) - 1 > \sum_{j = 1}^{p} R_j$ \\

When this relationship holds, we call it ``convexity".  This is the result of a sequence of repeated gains (or losses) that compound to a return that is greater than the arithmetic return.  The convexity dynamic can also be understood as repeatedly ``buying high" and increasing your exposure during a sequence of positive returns, or repeatedly ``selling low" and decreasing your exposure during a sequence of negative returns \cite{Tuzun2013}. Let us change our example to be 2 consecutive days of 1\% gains or 1\% losses ... \\

$((1 + 0.01) * (1 + 0.01)) - 1 = 0.0201 > 0.02 = 0.01 + 0.01$ \\

$((1 - 0.01) * (1 - 0.01)) - 1 = -0.0199 > -0.02 = -0.01 + -0.01$ \\

For longer time series, for convexity to be observed, it is not required that all returns are gains (or losses), only that sufficiently many of them are the same sign so that the cross products of the polynomial expansion on the left hand side are positive, and thus our inequality for the dynamic of convexity would hold.  Now we will extend these two conditions so that they are more suitable when used in the context of LETFs.  Let $D$ be the following function ...
\begin{flalign*}
D({\bf R}|\beta) &= \big[ \prod_{j = 1}^{p} (1 + \beta R_{j}) - 1 \big] - \big[ \beta(\prod_{j = 1}^{p} (1 + R_{j}) - 1) \big] &
\end{flalign*}

The first term is the $p$ day return produced by leveraged daily returns--what is achieved with the common LETF structure ignoring fees and tracking error.  The second term is the leveraged period return--what is achieved if an investor leveraged an initial capital investment in an index by a factor of $beta$, and ignored margin (borrowing) costs and tracking error. Our rationale for why the second term is not the leveraged arithmetic return $\beta \sum R_j$ will be clarified after the following derivation.
\begin{flalign*}
D({\bf R}|\beta) &= \big[ \prod_{j = 1}^{p} (1 + \beta R_{j}) - 1 \big] - \big[ \beta(\prod_{j = 1}^{p} (1 + R_{j}) - 1) \big] &\\
&= \big[ \beta \sum_{ j \in S_1 } R_{j} + \beta^2 \sum_{ j,k \in S_2 } R_jR_k + \beta^3 \sum_{ j,k,l \in S_3 } R_jR_kR_l + ... \big] - \big[ \beta \sum_{ j \in S_1 } R_{j} + \beta \sum_{ j,k \in S_2 } R_iR_j + \beta \sum_{ j,k,l \in S_3 } R_iR_jR_k + ... \big] &
\end{flalign*}

$S_1 = \{ 1, 2, ..., p \}$

$S_2 = \{ (a,b): a \in S_1, b \in S_1, a \ne b \}$

$S_3 = \{ (a,b,c): a \in S_1, b \in S_1, c \in S_1, a \ne b, b \ne c, a \ne c \}$

\begin{flalign*}
D({\bf R}|\beta) &= (\beta^2 - \beta) \sum_{ j,k \in S_2 } R_jR_k + (\beta^3 - \beta) \sum_{ j,k,l \in S_3 } R_jR_kR_l + ... &\\
& = \sum_{g = 2}^{p} (\beta^g - \beta) \big[ \sum_{h = 1}^{{p \choose g}} ( \prod_{j = 1}^{p} R_j^{I_A(p, g, h, j)} ) \big] &
\end{flalign*}

$I_A(p, g, h, j)$ indicator function for set $A$
\begin{flalign*}
A = \{ &(2,2,1,1), (2,2,1,2), &\\
        &(3,2,1,1), (3,2,1,2), (3,2,2,1), (3,2,2,3), (3,2,3,2), (3,2,3,3), (3,3,1,1), (3,3,1,2), (3,3,1,3), &\\
        & ... \} &
\end{flalign*}

Note the coefficient on the inner summation containing $\beta$.  If we were using the leveraged arithmetic return $\beta \sum_{j=1}^{p}R_j$ in our derivation, the $\beta$ term would instead be only $\beta^g$.  In the traditional case when there is no change in leverage the interpretation of the volatility drag condition that is based on the arithmetic return must be used.  Clearly, setting $\beta = 1$ makes the coefficient terms in our interpretation all zeros, thus the final value of $D$ would always be a noninformative zero. \\

But in the presence of leverage ($|\beta| \geq 1$), we prefer our interpretation because the arithmetic return has no real world significance.  In our function $D$, we use $\beta (\prod_{j=1}^{p} (1 + R_j) - 1)$ instead because this is an achievable return, where an investor earns the compound return for the period on a capital base that was leveraged $\beta$ times.  We feel this is a more meaningful calculation for determining volatility drag versus convexity.  Also, if instead the observed values for the individual $R_j$ variables were replaced with the geometric mean $\bar{R}_g = \prod_{j=1}^{p} (1 + R_j)^{1/p} - 1$, a leveraged arithmetic return would not remain the same while our leveraged period return would produce the same value.  This last reason for our preference will soon be a point of emphasis. \\

With our function $D$ properly defined, we state our revised criteria for determining volatility drag versus convexity when observing a series of leveraged returns (i.e. assuming $|\beta| > 1$) ... \\

$D({\bf R}|\beta) < 0 \implies$ volatility drag \\

$D({\bf R}|\beta) > 0 \implies$ convexity \\

We must note that $D$ is technically bounded below at -1.  In other words, the effect of volatility drag cannot be in excess of -100\%, because we cannot lose more than 100\% of our investment value.  Mathematically, this would be expressed by censoring the individual $(1 + \beta R_j)$ terms and the entire $\beta (\prod_{j=1}^{p} (1 + R_j) - 1)$ term at 0.  This lower bound still holds when $\beta < -1$ in our analysis of LETFs, because funds with a negative leverage multiple are still purchased assets with a Net Asset Value (NAV) and a share price, as opposed to shorting a stock where positions are initiated with sell transactions and losses have no theoretical lower bound. \\

There is also a conditional upper bound to our function $D$ dependent on the compound period return that is theoretical and not driven by real world constraints. This is because convexity is maximized when the daily return for all $p$ days is the geometric mean for the period.  We have ... \\

$R = \prod_{j=1}^{p} (1 + R_j) - 1$ \\

$\bar{R}_g = (1 + R)^{1/p} - 1$ \\

$(1 + \beta \bar{R}_g)^{p} - 1 \geq \prod_{j = 1}^{p} (1 + \beta R_{j}) - 1$ \\

$\big[(1 + \beta \bar{R}_g)^{p} - 1 \big] - \big[ \beta R \big] \geq \big[\prod_{j = 1}^{p} (1 + \beta R_{j}) - 1 \big] - \big[ \beta R \big]$ \\

$\big[(1 + \beta \bar{R}_g)^{p} - 1 \big] - \big[ \beta ((1 + \bar{R}_g)^{p} - 1) \big] \geq \big[\prod_{j = 1}^{p} (1 + \beta R_{j}) - 1 \big] - \big[ \beta (\prod_{j=1}^{p} (1 + R_j) - 1) \big]$ \\

$D(\bar{R}_g {\bf 1}_p|\beta) \geq D({\bf R}|\beta)$ \\

$D(\bar{R}_g {\bf 1}_p|\beta) = \sum_{j=2}^{p} (\beta^j - \beta) {p \choose j} \bar{R}^j_g \geq 0$ \\

$\forall {\bf R}, \beta$ s.t. $R_j \geq -1, |\beta| \geq 1$ \\

It is very important to note that both input vectors $\bar{R}_g {\bf 1}_p$ and ${\bf R}$ provide the same compound period return $R = \prod_{j=1}^{p} (1 + R_j) - 1$, but clearly represent different ways to arrive at that return.  This implies that regardless of the observed values of $R_j$ that compound to the observed value of $R$, we can calculate the maximum convexity that is possible with a leveraged, daily compounded fund structure knowing only $R$ (and the parameters $p$ and $\beta$).  Not coincidentally, this condition that maximizes convexity is equivalent to saying there was no volatility during the period (the return for each day is a constant $\bar{R}_g$, i.e. no randomness).  This point will become useful when we define our new volatility statistic in the next section. \\

\begin{knitrout}
\definecolor{shadecolor}{rgb}{0.969, 0.969, 0.969}\color{fgcolor}\begin{figure}
\includegraphics[width=\maxwidth]{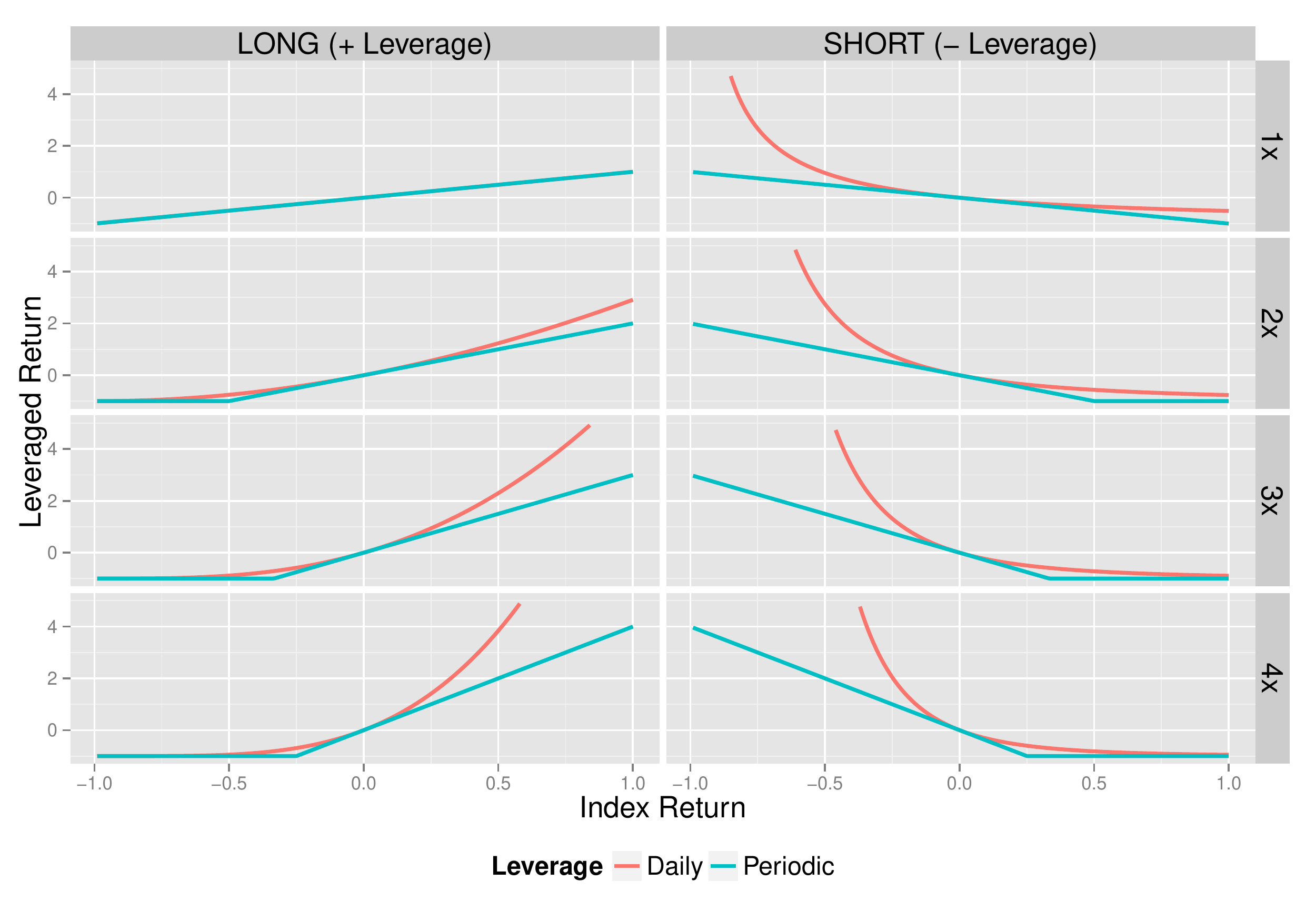} \caption[Leveraged Daily vs]{Leveraged Daily vs. Leveraged Periodic Returns (21 Days)}\label{fig:fig_vislev_21}
\end{figure}

\end{knitrout}
\clearpage

In Figure \ref{fig:fig_vislev_21} we visualize the return profile of leveraged returns, with and without daily compounding for various leverage multiples, which are indicated along the right edge of the plots.  The column labeled ``SHORT" plots returns where the sign on the leverage multiple is negative, and the ``LONG" column is for positive multiples.  The parameter $p$ is fixed at 21 days.  Each function represents a return profile for a hypothetical leveraged instrument, given the $x$ value as the return of the hypothetical underlying index over a 21 day period ... \\

$x$ := $\prod_{j=1}^{21} (1 + R_j) - 1$, given \\

The functions marked as periodic leverage ($y_{periodic}$, colored green) are simply this $x$ value multiplied by the indicated leverage multiple $\beta$ (and censored at -1, as we cannot lose more than 100\%).  For the functions marked as daily leverage ($y_{daily}$, colored red), we compound the geometric mean return implied by the value of $x$ (leveraged $\beta$ times) over 21 days ... \\

$y_{periodic} = \beta x$ \\

$y_{daily} = (1 + \beta \bar{R}_g)^{21} - 1$ \\

$\bar{R}_g = (1 + x)^{1/21} - 1$ \\

$y_{daily}$ is the maximum convexity curve given $x$.  From the figure, we see the curvature of the $y_{daily}$ curves increase as the magnitude of $\beta$ increases.  For positive values of $x$, curves defined by larger (in magnitude) leverage multiples dominate, and for negative values of $x$, curves defined by smaller leverage multiples dominate.  More generally, for a given series of returns, both volatility drag and convexity intensify as the magnitude of the leverage multiple increases.  This might be better explained with a reference to the $(\beta^g - \beta)$ term in the function $D$ defined earlier.  If $D({\bf R}|\beta) > 0$, then increasing $\beta$ will increase the value of $D$, and we say convexity has intensified.  If $D({\bf R}|\beta) < 0$, then increasing $\beta$ will decrease the value of $D$, and we say volatility drag has intensified. \\

Now that we have a better understanding of the choice of $\beta$ on the maximum convexity curve, let us consider the impact of $p$, the length of the returns vector ${\bf R}$ (or the number of days in the sample period).  Similar to increasing $\beta$, the curvature also increases as $p$ increases, given that the compound return $R$ is held constant.  However, the result of this change in $p$ is bounded as $p$ goes to infinity.  We have ... \\

$R = \prod_{j = 1}^{p} (1 + R_{j}) - 1$ \\

$\bar{R}_g = (1 + R)^{1/p} - 1$ \\

$\lim_{p \to \infty} (1 + \beta \bar{R}_g)^p - 1 = \lim_{p \to \infty} (1 + \beta ([1 + R]^{\frac{1}{p}} - 1))^p - 1 = (1 + R)^{\beta} - 1$ \\ 

Figure \ref{fig:fig_vislevmax} illustrates this result of taking the limit of $p$ days to infinity for the maximum convexity curve for a given leverage multiple, across various time horizons and  leverage multiples.  We see that curves defined by greater values of $p$ (the geometric mean return implied by $p$ is compounded with leverage over $p$ days) dominate for each choice of leverage multiple (across both positive and negative values of $x$). \\

$x$ := $\prod_{j=1}^{21} (1 + R_j) - 1$, given \\

$y_{periodic} = \beta x$ \\

$y_{daily, 5} = (1 + \beta \bar{R}_g)^{5} - 1 = (1 + \beta ((1 + x)^{1/5} - 1))^{5} - 1$ \\

$y_{daily, 10} = (1 + \beta \bar{R}_g)^{10} - 1 = (1 + \beta ((1 + x)^{1/10} - 1))^{10} - 1$ \\

$y_{daily, \infty} = \lim_{p \to \infty} (1 + \beta \bar{R}_g)^{p} - 1 = (1 + R)^{\beta} - 1$ \\

$\bar{R}_g = (1 + x)^{1/p} - 1$ \\

\begin{knitrout}
\definecolor{shadecolor}{rgb}{0.969, 0.969, 0.969}\color{fgcolor}\begin{figure}
\includegraphics[width=\maxwidth]{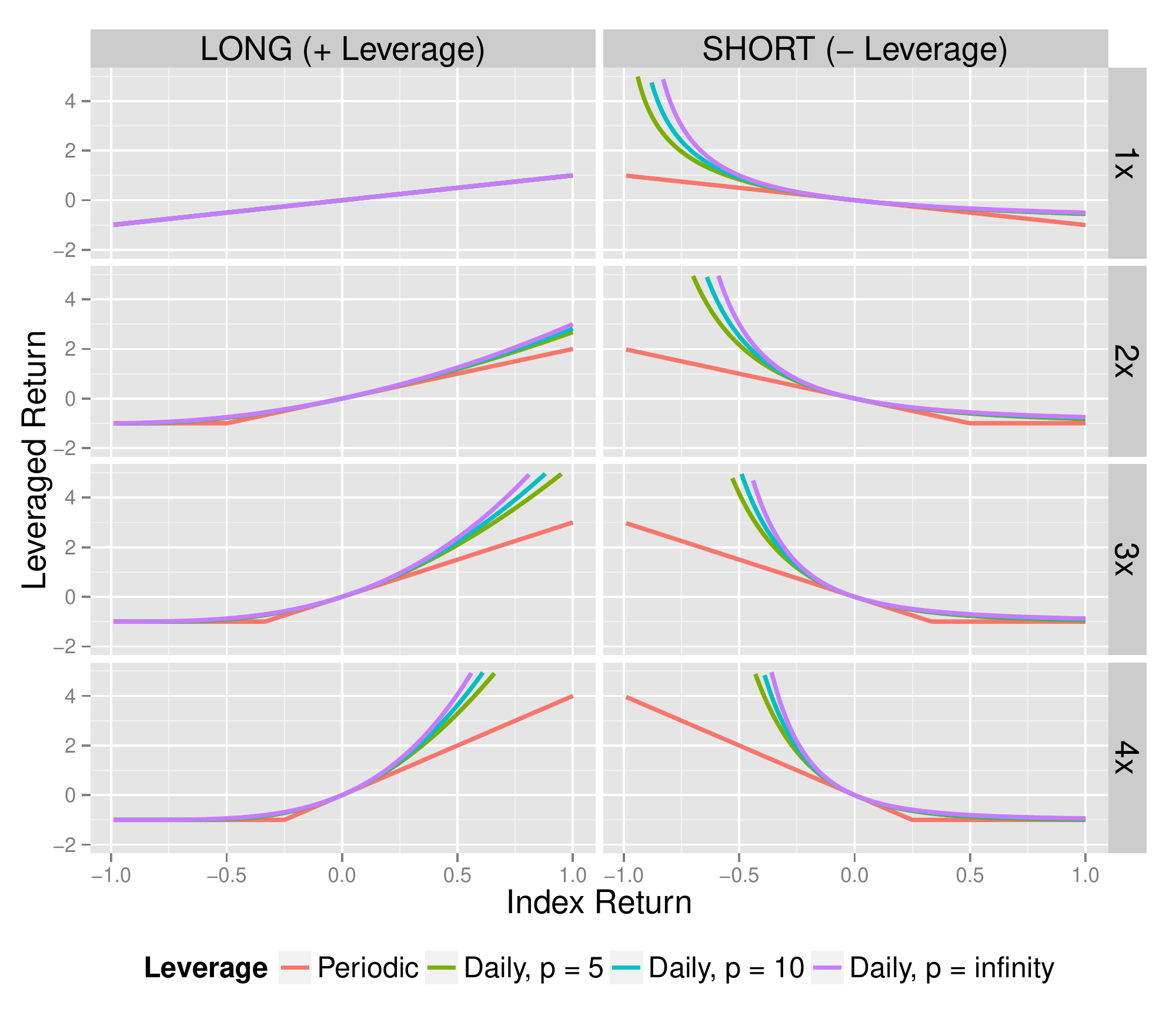} \caption[Leveraged Daily vs]{Leveraged Daily vs. Leveraged Periodic Returns (Various values for p days)}\label{fig:fig_vislevmax}
\end{figure}

\end{knitrout}

\clearpage


\section{Shortfall from Maximum Convexity} \label{chapter_smc}


Going all the way back to the Markowitz' benchmark paper on portfolio theory, it has been commonly accepted to model the volatility of asset returns as the standard deviation of the returns distribution \cite{Markowitz1952}.  Accordingly, sample standard deviation is used to measure the realized volatility of a $p$ day sample of returns ($s = [\frac{1}{p-1} \sum_{j=1}^{p} (R_j - \bar{R})^2]^{\frac{1}{2}}$).  In this chapter, we propose an alternative statistic for measuring the realized volatility of LETF returns specifically.  Our statistic is based on the idea of volatility drag, and we feel it has an interpretation that is more intuitive than that of standard deviation and because it allows for greater variability (when compared to standard deviation on the same dataset), we feel that it is more informative as well.  It is a hybrid summary of returns and volatility where the values are returns but their interpretation is to quantify the level of risk associated with the returns.  \\


The major criticism of sample standard deviation and a motivation for our approach is that the assumptions of independence (and identical distribution) and Normality required for interpreting results do not hold with respect to returns data.  Consider the following derivation where we model daily, log returns so that the $p$ day return is additive ...\\

$R_1, R_2, ..., R_p$ daily returns, $X_j = \log(1 + R_j)$ log, daily returns \\

$X_1, X_2, ..., X_p \sim$ Normal i.i.d. \\

$Y = \sum_{j=1}^{p} X_j$ log, $p$ day return \\\

$Var[Y] = p * Var[X]$ \\

$s_{Y} = \sqrt{p} * s_X = \sqrt{p} \sqrt{ \frac{1}{p-1} \sum_{j=1}^{p} (x_j - \bar{x})^2} \approx \sqrt{ \sum_{j=1}^{p} (x_j - \bar{x})^2 }$ \\

If the assumption of i.i.d. Normal data holds, then $s_Y$ is our estimate of the standard deviation of the $p$ day log returns distribution, and we could say that roughly 68\% of observations will fall within $\pm s_Y$ of the mean.  However, it is now commonly understood that asset returns data are not i.i.d. Normal.  We provide the Normal Quantile-Quantile plots for the returns series of LETF ticker SPXL (the +3x LETF tracking the S\&P500) in Figure \ref{fig:fig_qqplots} as our evidence against a Normal distribution.  Many of the sample quantiles of the returns data, especially noting the quantiles corresponding to the tails of the distributions, do not match the corresponding quantiles of a Normal distribution.  Figure \ref{fig:fig_qqplots} includes the ACF and PACF plots of the same data, which we present as evidence that the data is not independent.  Note that lags 1 and 5 break the red lines for the 95\% confidence interval in both the ACF and PACF plots.  \\

If we accept this evidence, then we must revise the derivation above to account for some non-Normal returns distribution and the lack of independence.  We will allow for the assumption that the daily data are identically distributed, however.  After these changes, we have ... \\

$X_1, X_2, ..., X_p \sim f_X$ identically distributed, but dependent \\

$Y = \sum_{j=1}^{p} X_j$ log, $p$ day return \\\

$Var[Y] = p * Var[X] + \sum_{i \neq j} Cov[X_i, X_j]$ \\

$s_{Y} = \sqrt{p * s_X^2 + \sum_{i \ne j} s_{X_i, X_j}^2} \ne \sqrt{p} * s_X$ \\

If the true process that is said to generate LETF returns is not Normal and does not generate data independently (or at least without correlation), then the interpretation of the quantity $\sqrt{p} * s_X$ is nearly meaningless.  It is not an estimator of the standard deviation of the $p$ day log return $Y$, and even if we did calculate a proper estimate we can no longer claim that 68.27\% of observations will fall within $\pm 1$ standard deviation of the mean.  We are left with a nonparametric interpretation that is based on the literal formula for the sample standard deviation estimator--the 2-norm of the centered observation vector of log returns ($\sqrt{p} * s_X = ||{\bf x} - \bar{x}||_2$).  By nature of the squaring this statistic does not produce any sort of intuitive, real-world quantity.  By contrast, we consider Maximum Drawdown, another commonly used statistic, as providing an intuitive, real-world quantity (a rate of return) that is immediately understood when trying to communicate risk.  Without the necessary assumptions that make $p * s_X^2$ a true estimator of the population variance for some Normal distribution of returns, and without a strong intuition for interpreting the 2-norm of returns samples absent any theoretical foundation, our goal is to create an improved statistic for measuring the volatility of LETF returns.  \\

\clearpage
\begin{knitrout}
\definecolor{shadecolor}{rgb}{0.969, 0.969, 0.969}\color{fgcolor}\begin{figure}

{\centering \includegraphics[width=\maxwidth]{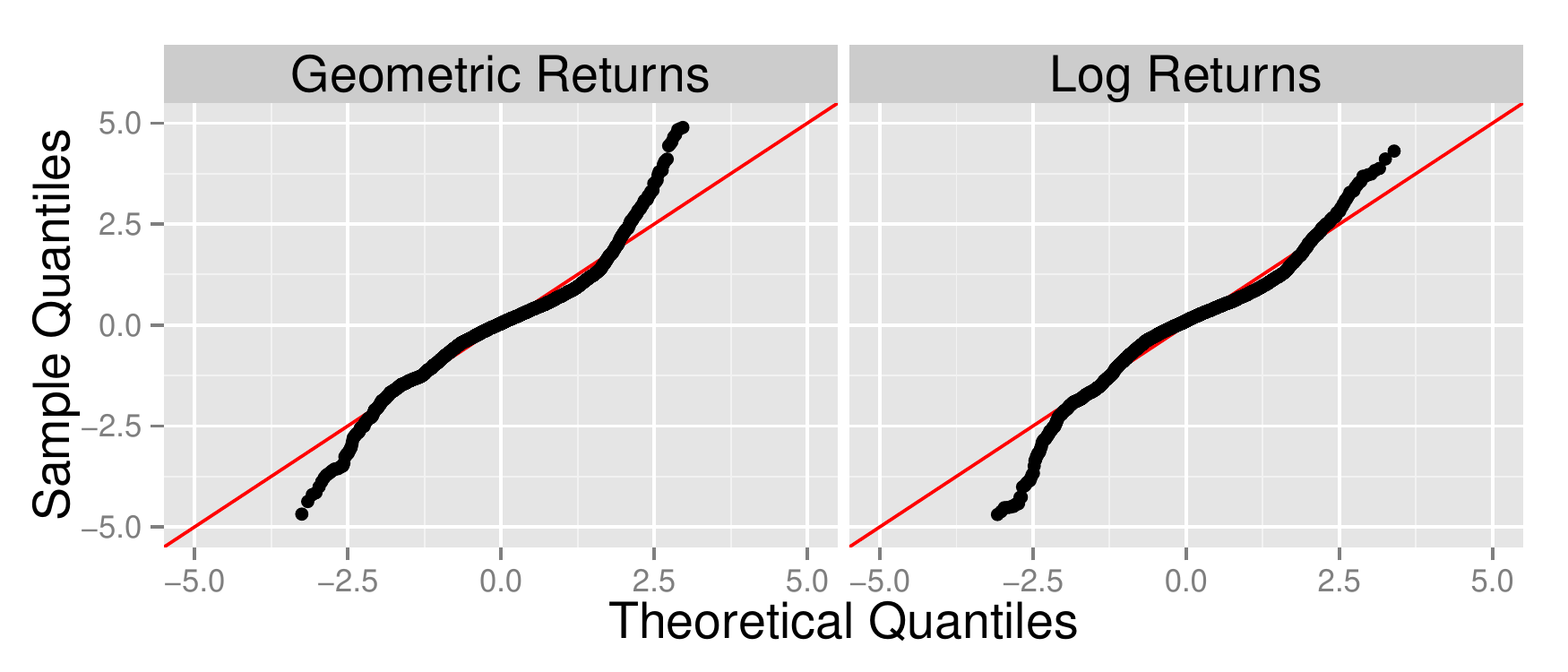} 

}

\caption[Quantile-Quantile Plots for Daily SPXL Returns]{Quantile-Quantile Plots for Daily SPXL Returns}\label{fig:fig_qqplots}
\end{figure}

\end{knitrout}

\begin{knitrout}
\definecolor{shadecolor}{rgb}{0.969, 0.969, 0.969}\color{fgcolor}\begin{figure}

{\centering \includegraphics[width=\maxwidth]{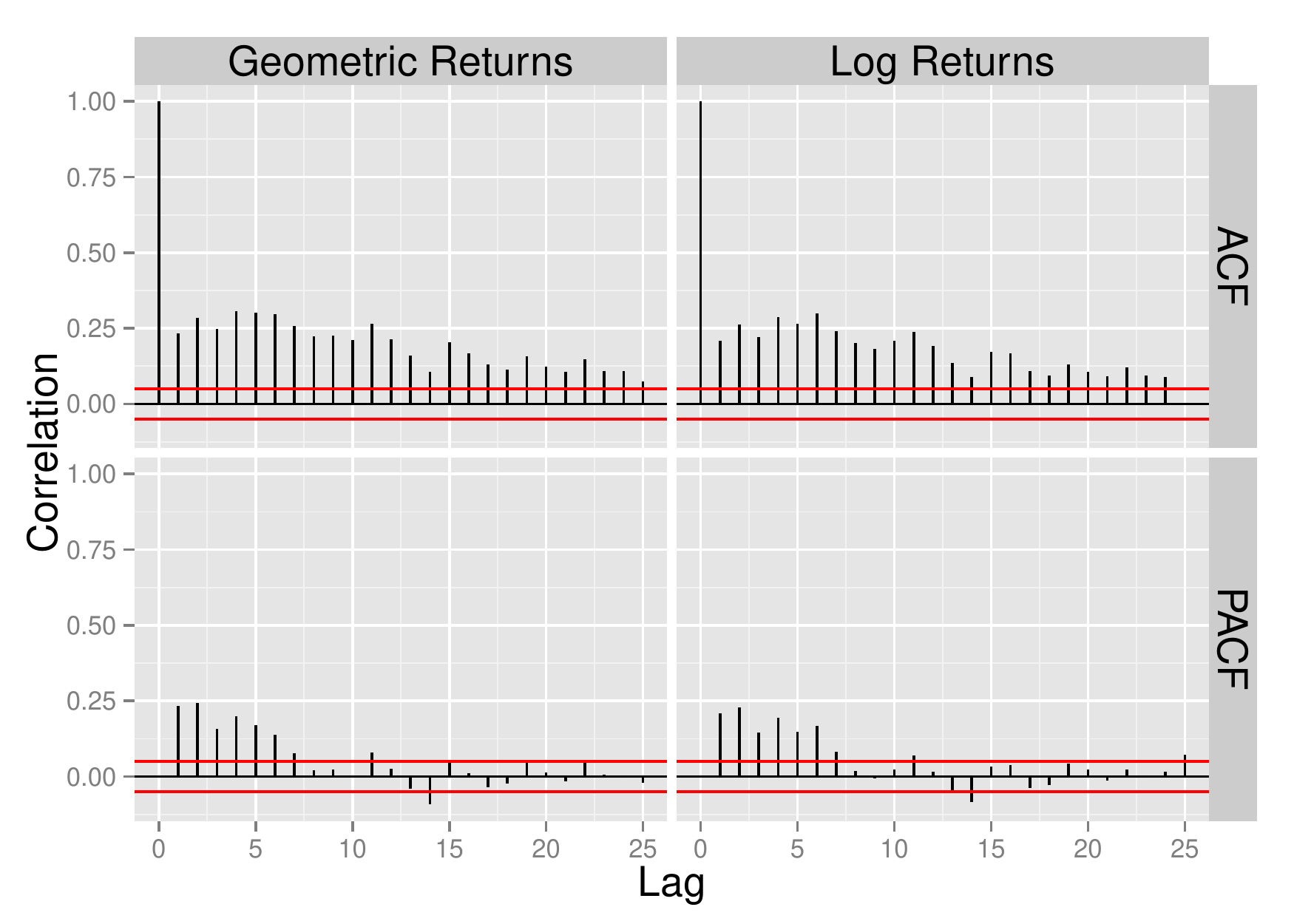} 

}

\caption[ACF and PACF Plots for Squared Daily SPXL Returns]{ACF and PACF Plots for Squared Daily SPXL Returns}\label{fig:fig_acfplots}
\end{figure}

\end{knitrout}
\clearpage

\subsection{Define Shortfall from Maximum Convexity}


Let us review the concept of maximum convexity that was defined in the previous section and use it as the foundation for defining our new statistic. \\

For a given compound return $R$ on an asset over a period of $p$ days, the hypothetical leveraged, daily compounded return derived from that asset's daily returns $R_j$s has an upper bound, which we shall denote as $R_{MAX}$.  This return is achieved if the geometric mean of the return series is observed every day, such that the value is dependent on the choice of leverage multiple $\beta$, but the geometric mean of the asset's return series is not. \\

For a given leverage multiple $\beta$, and a series of $p$ daily total returns on an index $R_1, R_2, ..., R_p$, we have ... \\

$R_{Index} = \prod_{j=1}^{p} (1 + R_{j}) - 1$ \\

$\bar{R}_{Index} = (1 + R_{Index})^\frac{1}{p} - 1$ \\

$R_{MAX} = (1 + \beta \bar{R}_{Index})^p - 1$ \\

We say that $R_{MAX}$ is the maximum return that could be achieved by a series of $p$ daily returns that compounds to $R_{Index}$ (given the leverage multiple), and so it also maximizes the degree of convexity as was defined in the previous section ... \\

$R_{MAX} = (1 + \beta \bar{R}_{Index})^p - 1 \geq \prod_{j=1}^{p} (1 + \beta X_{j}) - 1$, $\forall {\bf X}$ s.t. $\prod_{j=1}^{p} (1 + X_{j}) - 1 = R_{Index}$ \\

$\implies$ $R_{MAX} - \beta R_{Index} \geq [\prod_{j=1}^{p} (1 + \beta X_{j}) - 1] - \beta R_{Index}$ \\

The interpretation of maximum convexity is that the effect of compounding gains on gains (or losses on losses) over the $p$ day period could produce a return up to $R_{MAX}$, which by definition is greater than or equal to $\beta R_{Index}$ (the result of earning the index return on a leveraged capital base).  \\

We are now ready to define our volatility statistic that was designed specifically for LETFs, Shortfall from Maximum Convexity, which we shall abbreviate as SMC.  This new statistic is defined to be the geometric excess return of the maximum return $R_{MAX}$ with respect to the LETF return observed over the $p$ day period. \\

For a series of $p$ daily returns on an LETF, $R_{LETF,1}, R_{LETF,2}, ..., R_{LETF,p}$, where the LETF has a stated leverage multiple of $\beta$, and the series of $p$ daily total returns on the corresponding underlying index $R_1, R_2, ..., R_p$, we have ... \\


$R_{LETF} = \prod_{j=1}^{p} (1 + R_{LETF,j}) - 1$ \\

$R_{MAX} = (1 + SMC)(1 + R_{LETF}) - 1$ \\

$SMC({\bf R}_{LETF}, {\bf R}_{Index}|\beta) = (1 + R_{MAX}) / (1 + R_{LETF}) - 1$ \\

Where $R_{MAX}$ is defined as above using the given index returns along with the stated leverage multiple for the LETF. \\

SMC values are interpreted as the performance shortfall from the hypothetical return $R_{MAX}$.  To help demonstrate, we have included Figure \ref{fig:fig_visdef}, a visual example of calculating SMC for a single 252 day period (tick marks on axes and stated statistics are in decimal form). The single green point is a hypothetical 252 day observation when a +3x LETF returned 50\% and the underlying index returned 35\%.  The blue curve represents $R_{MAX}$ at the given index return, and the purple line represents the given index return multiplied by $+3$ (and bounded at -1, or -100\%).  For our hypothetical observation with an index return of 35\%, the corresponding point on the $R_{MAX}$ curve is 145.77\%, which results in a final SMC value of 63.85\%.  Ignoring future movement of the underlying index, the LETF would have to return 63.85\% to match the maximum return that was achievable for this 252 day period. \\

\begin{knitrout}
\definecolor{shadecolor}{rgb}{0.969, 0.969, 0.969}\color{fgcolor}\begin{figure}

{\centering \includegraphics[width=\maxwidth]{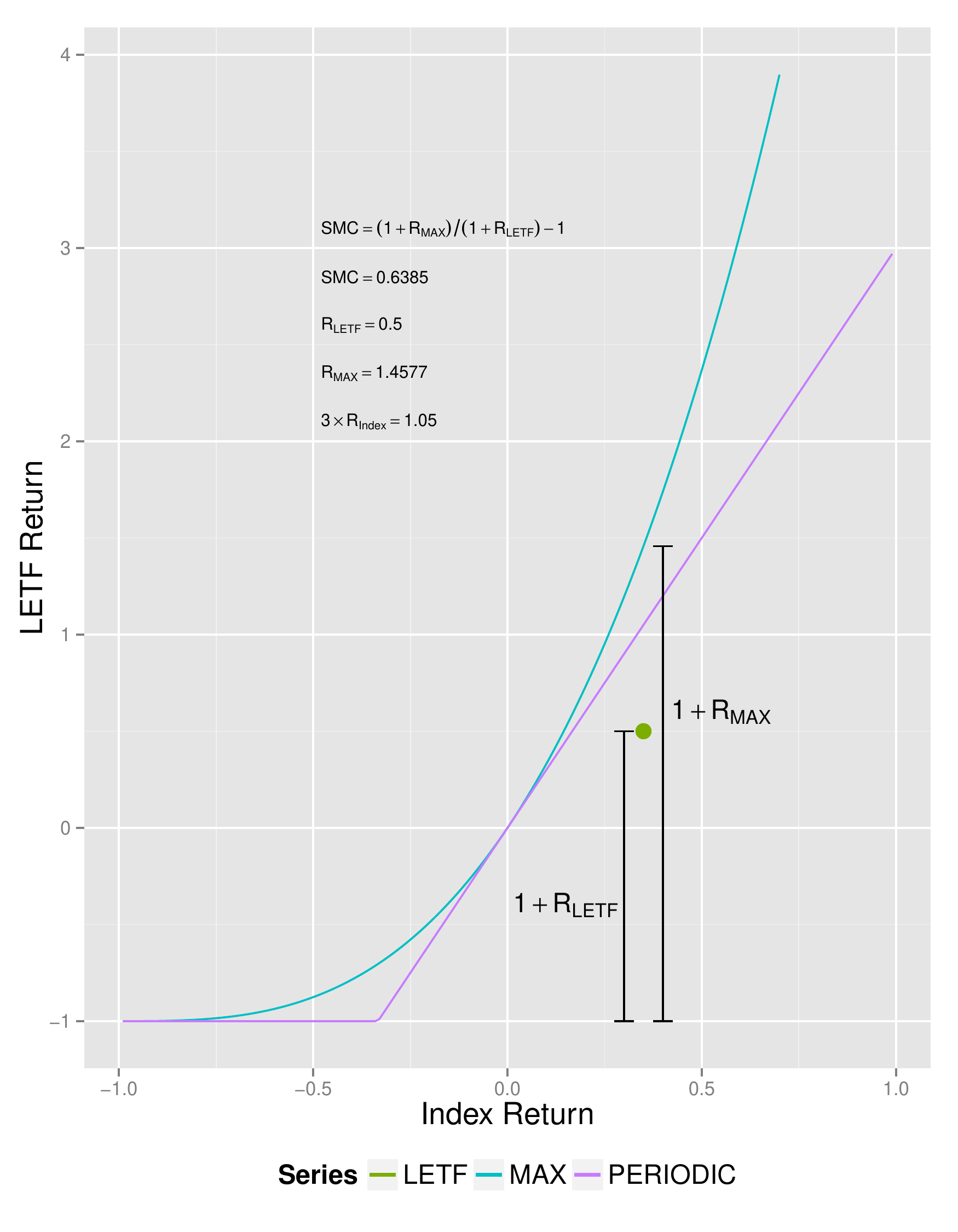} 

}

\caption[Index Returns vs]{Index Returns vs. LETF Returns for Hypothetical Index and +3x LETF over 252 Day Periods}\label{fig:fig_visdef}
\end{figure}

\end{knitrout}
\clearpage

\begin{knitrout}
\definecolor{shadecolor}{rgb}{0.969, 0.969, 0.969}\color{fgcolor}\begin{figure}

{\centering \includegraphics[width=\maxwidth]{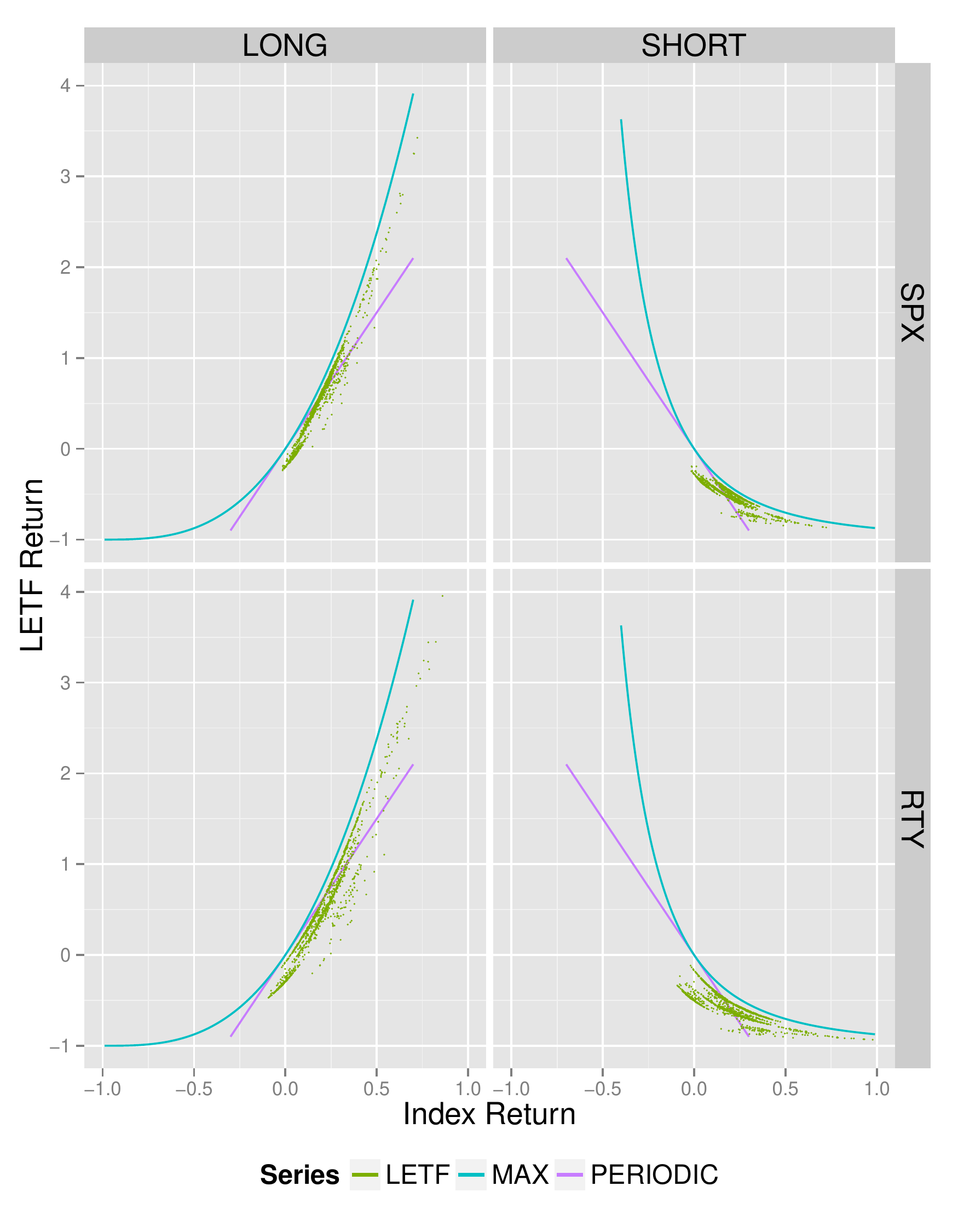} 

}

\caption[Index Returns vs]{Index Returns vs. LETF Returns for LETF Ticker Pairs on SPX, RTY over 252 Day Periods}\label{fig:fig_vishist_smc_demo}
\end{figure}

\end{knitrout}
\clearpage


Figure \ref{fig:fig_vishist_smc_demo} is essentially the same as Figure \ref{fig:fig_visdef}, but using real data from the $\pm$3x LETF ticker pairs with underlying indexes SPX and RTY (S\&P500 and Russell 2000) over 252 day periods.  We make a special note that visuals such as this one were instrumental in our LETF analysis and played a significant role in defining SMC.  The specific choices of indexes are not important, only that we selected a pair where it is commonly accepted that one is riskier than the other.  In our case, we view the S\&P500 as less risky than the Russell 2000.  Each green point marks an observed return for the specified LETF (as indicated by index ticker and side) and number of days, and as before the blue curve represents $R_{MAX}$ at a given index return and the purple line represents the index return multiplied by $+3$ (long side) or $-3$ (short side).  The key observation is that the green points seem to disperse away from the blue curve as we consider an LETF (on the same side) based on a riskier index.  Our intuition was that greater dispersion is indicative of greater volatility, and SMC was defined specifically to measure that dispersion.  \\

Our definition of SMC is not yet final as we must properly account for sample periods when the observed LETF return is greater than $R_{MAX}$.  We will observe this result in the repeated presence of abnormally large, positive tracking errors.  In our data set containing 20 LETFs this was observed in about 1\% of all 252 day periods.  Because of this possibility, we have ... \\

$SMC({\bf R}_{LETF}, {\bf R}_{Index}|\beta) = max\{0, (1 + R_{MAX}) / (1 + R_{LETF}) - 1\}$ \\

Alternatively, we could add the maximum observed tracking error to the $\beta \bar{R}_{Index}$ term before calculating $R_{MAX} = (1 + \beta \bar{R}_{Index})^p - 1$, or simply allow for negative values of SMC.  These options would generate results that are different than what is presented at the end of this section (average SMC values would increase slightly if incorporating the maximum tracking error, or decrease slightly if allowing negative values), otherwise applications of our new statistic would remain unchanged.  We choose to censor at 0 because it is simpler than computing a maximum tracking error and lets us avoid any confusion that may result from interpreting negative values.  \\

SMC not only captures the impact of volatility drag produced by the daily compounding of leveraged returns, but also the impact produced by the compounding of tracking errors and fees.  We demonstrate this and the dynamic nature of the relationship between $R_{MAX}$ and $R_{LETF}$ with 2 examples of calculating SMC over 252 day periods in Figure \ref{fig:fig_smcex}.  Following the color scheme from previous figures, the blue line represents $R_{MAX}$ at each day, the purple line is the leverage multiple times the index return up to that date, and the green line is the observed LETF return.  An orange line is also included, which is the underlying index return compounded daily at the given leverage multiple, or what the LETF return should be without fees or tracking errors.  \\

The scenario in the top panel is for the period ending 12/31/2013, for the +3x LETF ticker UDOW (underlyer Dow Jones Industrials).  We included this scenario to demonstrate the advantage of leveraged, daily compounding of returns experienced in a low volatility environment.  The green line for the observed return on UDOW finishes just below the blue line for $R_{MAX}$, and we observe a low SMC of 0.0515.  This means that UDOW came close to achieving the maximum potential return over the period given the leverage multiple of +3 and the performance of the index, and because TZA outperformed the 3 times the Dow Jones return, we have observed the convexity dynamic.  By the end of the period we see some separation between the orange line (theoretical LETF without fees or tracking errors) and green line, which tells us that fees and tracking error were partially responsible for the performance shortfall.  \\

The scenario in the bottom panel is for the period ending 11/30/2009, for the -3x LETF ticker TZA (underlyer Russell 2000).  This scenario is included to demonstrate the disadvantage of leveraged, daily compounding of returns experienced in a high volatility environment.  Here, the green line finishes further below the blue line than in the first scenario, and we observe a high SMC of 1.8651.  Because the purple line (-3 times the Russell 2000 performance) ends the period above the green line, we have observed the dynamic of volatility drag.  TZA vastly underperformed the maximum potential return, and because of the minimal separation between orange line and green line (observed TZA performance), we can say that the shortfall was caused by the volatility drag associated with daily, leveraged compounding and not necessarily the accumulated drag associated with fees or tracking error. \\

\begin{knitrout}
\definecolor{shadecolor}{rgb}{0.969, 0.969, 0.969}\color{fgcolor}\begin{figure}

{\centering \includegraphics[width=\maxwidth]{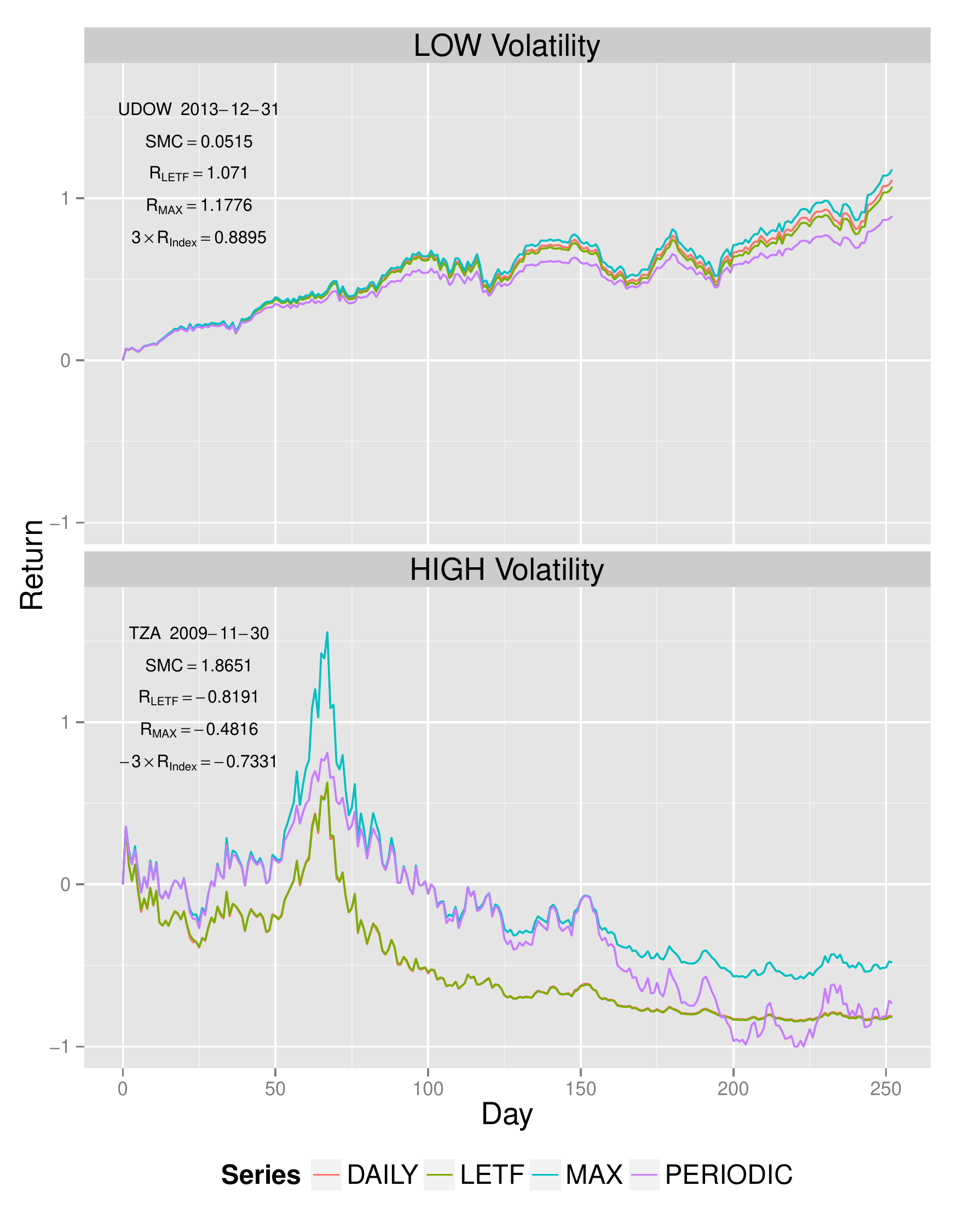} 

}

\caption[Various Return Series over 252 Day Periods of High and Low Volatility]{Various Return Series over 252 Day Periods of High and Low Volatility}\label{fig:fig_smcex}
\end{figure}

\end{knitrout}
\clearpage


We close our introduction of SMC with a discussion on the perceived asymmetry of results when using the statistic with a long and short LETF pair over the same period.  At first glance, it may appear as though the statistic is biased to produce greater values with short side LETF returns data ($\beta < -1$).  It occurs in the overwhelming majority of periods available in our data set, but we show this does not have to be true because of the counter examples that do exist. \\

We are specifically referring to the following relationship, where we consider a pair of long and short LETFs on the same underlying index with equal but opposite leverage multiples, and no tracking errors or fees ... \\

$SMC(\beta{\bf R}, {\bf R} |\beta) \stackrel{?}{=} SMC(-\beta{\bf R}, {\bf R}|$$-{\beta})$ \\

${\bf R} = \begin{bmatrix} R_1 & R_2 & \cdots & R_p \end{bmatrix}$, vector of $p$ daily total returns on an index \\

We initially attempted a proof based on Jensen's inequality showing that short side SMC is guaranteed to be greater.  The set up for that proof proceeds as follows: for a given leverage multiple $\beta$, and a series of $p$ daily total returns on an index $R_1, R_2, ..., R_p$, we have ... \\

$R = \prod_{j=1}^{p} (1 + R_{j}) - 1$ \\

$\bar{R} = (1 + R)^\frac{1}{p} - 1$ \\

and require that $-1 \leq \beta R_j$ and $-1 \leq -\beta R_j$, because these terms represent LETF returns (without fees or tracking errors), and cannot be less than -100\%.  Also, the basis for our inequality is the result of changing the sign of $\beta$, so assume $\beta \geq 1$ (and $-\beta \leq -1$).  Our attempted proof continues ... \\

$SMC(\beta{\bf R}, {\bf R} |\beta) \stackrel{?}{=} SMC(-\beta{\bf R}, {\bf R}|$$-{\beta})$ \\

\begin{doublespace}

$\dfrac{(1 + \beta \bar{R})^p}{\prod_{j=1}^{p} (1 + \beta R_{j})} - 1 \stackrel{?}{=} \dfrac{(1 - \beta \bar{R})^p}{\prod_{j=1}^{p} (1 - \beta R_{j})} - 1$ \\

$\big[ \prod_{j=1}^{p} \dfrac{(1 - \beta R_{j})}{(1 + \beta R_{j})} \big]^{1/p} \stackrel{?}{=} \dfrac{(1 - \beta \bar{R})}{(1 + \beta \bar{R})}$

\end{doublespace}

Let $X_j = 1 + R_j$ to guarantee non-negativity with our terms, and note that $\bar{X}_g = \prod_{j=1}^{p} X_j^{1/p} = 1 + \bar{R}$.  With these substitutions, we can rearrange terms ... \\

\begin{doublespace}

$\big[ \prod_{j=1}^{p} \dfrac{(1 - \beta X_{j} + \beta)}{(1 + \beta X_{j} - \beta)} \big]^{1/p} \stackrel{?}{=} \dfrac{(1 - \beta \bar{X}_g + \beta)}{(1 + \beta \bar{X}_g - \beta)}$

\end{doublespace}

Now let $f(x) = (1 - \beta x + \beta) / (1 + \beta x - \beta)$, and re-write the inequality ... \\

$\prod_{j=1}^{p} f(X_j)^{1/p} \stackrel{?}{=} f(\bar{X}_g) = f(\prod_{j=1}^{p} X_j^{1/p})$ \\

On the domain of $f$ allowed by our constraints for $R_j$ and $\beta$, $f$ is a convex function, and thus Jensen's inequality provides the following ... \\

$\frac{1}{p} \sum_{j=1}^{p} f(X_j) \geq f(\frac{1}{p}\sum_{j=1}^{p} X_j)$ \\

Obviously, this statement does not help our cause.  If we could prove that it holds using geometric means opposed to arithmetic means, it would assert that long side SMC is always greater than or equal to short side SMC, which is not true.  Perhaps more importantly, we cannot prove that it holds with geometric means.  Given that $f$ is decreasing and the geometric mean of a non-negative random variable is less than or equal to the arithmetic mean, we can prove ... \\

$f(\prod_{j=1}^{p} X_j^{1/p}) \geq f(\frac{1}{p}\sum_{j=1}^{p} X_j)$ \\

$\frac{1}{p} \sum_{j=1}^{p} f(X_j) \geq \prod_{j=1}^{p} f(X_j)^{1/p}$ \\

Unfortunately, neither statement helps our proof, and a log transform does not make things easier, as the transformed right hand side would not permit the use of Jensen's inequality.  At this point, we reconsidered our pursuit of a theoretical proof of the inequality and exhaustively searched the data for counter examples. \\

Using the entirety of our data set of 10 underlying indexes and including leverage multiples of $\pm 1, \pm 2, \pm 3$, we have 95,913 sample periods of 21 days, and 93,603 sample periods of 252 days.  There were 25 instances (23 periods of 21 days, 2 periods of 252 days) when long side SMC was greater than the short side SMC (i.e. $SMC(\beta{\bf R}, {\bf R} |\beta) > SMC(-\beta{\bf R}, {\bf R}|$$-{\beta})$).  Within the 25 instances, there was no evidence of necessary or sufficient conditions that might help explain the direction of the inequality.  There were instances of both less than and greater than relationships in the presence of positive and negative index returns, positive and negative long and short LETF returns (i.e. $\beta R$ and $-\beta R$), and other boolean conditionals. \\

\begin{knitrout}
\definecolor{shadecolor}{rgb}{0.969, 0.969, 0.969}\color{fgcolor}\begin{figure}
{
\centering
\includegraphics[width=\maxwidth]{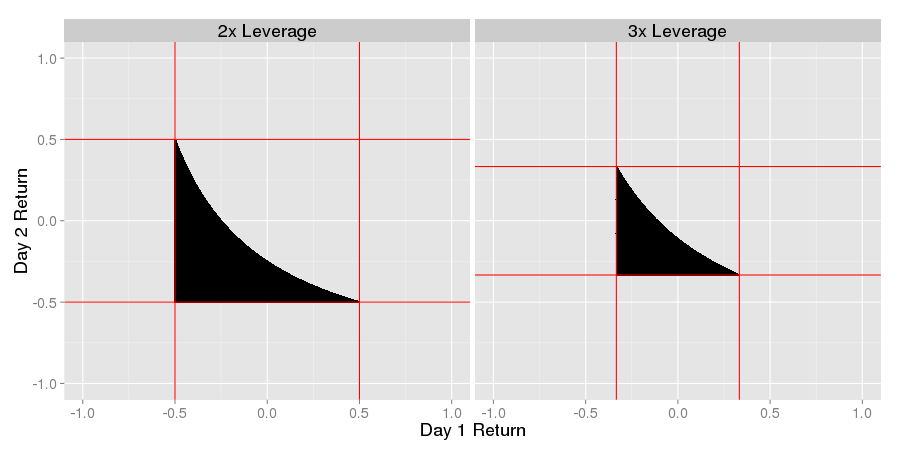}
}
\caption[Two Day Samples with Long Side SMC Greater Than Short Side SMC]{Two Day Samples with Long Side SMC Greater Than Short Side SMC}\label{fig:fig_visregion_longshort}
\end{figure}
\end{knitrout}

We can plot the region where long side SMC is greater than short side SMC, and we do this for all possible 2 day sample periods using $\beta = 2$ and $\beta = 3$ (``2x Leverage" and ``3x Leverage", respectively) in Figure \ref{fig:fig_visregion_longshort}.  The x axis represents all possible returns for the first day in the sample, and the y axis represents the return on the second day.  The red lines represent the effective boundary for daily returns allowed by the respective leverage multiples.  The shaded regions represent samples that would produce a long side SMC that is greater than the short side SMC, and equality occurs along the curved boundary of the region.  Technically, equality occurs along the main diagonal ($x = y$) as well, where 0 is observed on both the long and short side, but this detail is omitted for clarity.  By comparing the left and right subplots, we see that the curved boundary disperses from the origin as the leverage multiple decreases, such that there is no region that exists where long side SMC is greater than short side SMC with $\beta = 1$, because in that case long side SMC is always 0 by definition.  The curved boundary shifts closer towards the origin as the leverage multiple increases.  As the number of days in the sample increases beyond 2, we have an analogous multidimensional surface for the boundary of equality of long and short side SMC that continues to disperse further from the origin.  \\

Keeping these properties in mind, it becomes apparent as to why short side SMC is almost always greater in practice, given the nature of asset returns distributions.  For example, for 252 day periods and $\beta = 3$, we would need to observe a sample with one day experiencing a nearly -16\% return, and all other 251 days less than or equal to 0\%.  That kind of 1 day loss is an incredibly rare occurrence, and you would still need a strong negative trend across all other days in the sample period.  Ultimately, the definition of the SMC statistic does not guarantee greater values when using a negative leverage multiple, but the support of existing index returns distributions is such that there is very low probability mass over the regions where the converse is true. \\

%
%
%
%
%
%
%
%
%

\subsection{Fundamental Motivation for SMC}


We have properly defined our new statistic for measuring realized volatility and now we demonstrate how it is fundamentally different from the quantity $\sqrt{p} * s$, where $p$ is the chosen number of days in the sample period and $s$ is the maximum likelihood estimator (biased) of standard deviation.  Going forward we will refer to this quantity as the sample 2-norm, or SN2, because of our earlier proof where we demonstrated this quantity is not actually an estimator of the standard deviation of the $p$ day log return when using correlated data.  We will use log returns when calculating SN2 so that the $p$ day return is additive and the mean return represents the log of the geometric mean as opposed to arithmetic mean (although in practice not using log returns does not weaken our evidence presented later).  We have ... \\

$\bar{R}_{LETF} = \prod_{j=1}^{p} (1 + R_{LETF,j})^\frac{1}{p} - 1$ \\

$SN2({\bf R}_{LETF}) = \sqrt{p} * \sqrt{\frac{1}{p} \sum_{j=1}^{p} (\log(1 + R_{LETF,j}) - \log(1 + \bar{R}_{LETF}))^2}$ \\

$SN2({\bf R}_{LETF}) = \sqrt{\sum_{j=1}^{p} (\log(1 + R_{LETF,j}) - \log(1 + \bar{R}_{LETF}))^2}$ \\

Another motivating factor for our choice of using log returns when calculating SN2 can be seen in the following ... \\

$SN2({\bf R}_{LETF}) = \sqrt{\sum_{j=1}^{p} [ \log((1 + \bar{R}_{LETF}) / (1 + R_{LETF,j})) ]^2 }$ \\

$SMC({\bf R}_{LETF}, {\bf R}_{Index}|\beta) = \exp \big[ \sum_{j=1}^{p} [ \log((1 + \beta \bar{R}_{Index}) / (1 + R_{LETF,j})) ] \big] - 1$ \\

After rearranging terms, we see that SN2 is the 2-norm of a $p$ day sample of log, geometric excess returns.  SMC (without censoring) is the exponentiation of the sum of a $p$ day sample of log, geometric excess returns, but using $\beta \bar{R}_{Index}$ instead of $\bar{R}_{LETF}$ when computing the excess returns.  By using log returns when calculating SN2, we are also making a fair comparison with SMC in addition to the data transformation reasons mentioned earlier. \\

Both SMC and SN2 have a sample distribution that is bounded on the left at 0 and extends to $+\infty$ on the right.  Technically, the distribution for SMC includes $+\infty$, which is observed if the LETF return for the given sample period is $-100\%$.  For both statistics, a value of 0 is observed when the LETF achieves the sample mean return for every day in the period, with the interpretation being that there was no realized volatility observed (the underlying random process produced a constant return).  For SN2 the sample mean return is that of the LETF, but in the case of SMC, the sample mean return refers to the mean of the underlying index returns for the same sample period (multiplied by the stated leverage multiple of the LETF).  \\

We acknowledge the exception that must be made for the negative SMC values that are possible before censoring because of tracking error. For those samples, censoring SMC to 0 means that because of a beneficial sequence of tracking errors, there was no shortfall from the maximum return.  Also, we acknowledge that computing SMC requires the underlying index returns in addition to LETF returns.  \\

%

SN2 is asymmetric with respect to the leverage parameter $\beta$, but the difference is less than the difference that results when using the same inputs with SMC (and ignoring tracking error such that for daily returns $R_{LETF} = \beta R_{Index}$).  If we were not using log returns with SN2, it is trivial to check that output values are identical when using positive and negative leverage multiples ($[\sum_{j=1}^{p} (\beta R_j - \beta \bar{R})^2]^{\frac{1}{2}} = [\sum_{j=1}^{p} (-\beta R_j + \beta \bar{R})^2]^{\frac{1}{2}}$). With log returns, however, the daily returns are not equal ($\log(1 + \beta R_j) \ne \log(1 - \beta R_j)$), and there will be minor differences in the resulting values ... \\

$SN2(\beta {\bf R}) \leq SN2(-\beta {\bf R})$, $\forall {\bf R}$ s.t. $\bar{R} = \prod_{j=1}^{p} (1 + R_j)^\frac{1}{p} - 1 \geq 0$ \\

$SN2(\beta {\bf R}) \geq SN2(-\beta {\bf R})$, $\forall {\bf R}$ s.t. $\bar{R} = \prod_{j=1}^{p} (1 + R_j)^\frac{1}{p} - 1 \leq 0$ \\

SN2 is the 2-norm of a sample of centered, log returns, and due to the squaring and square root operations, has no real-world meaning.  Functionally, it measures the distance of the observed returns vector from the observed mean return.  SMC is the geometric excess return of the maximum return $R_{MAX}$ with respect to the observed LETF return.  It tells us the underperformance of the LETF relative to a hypothetical underlying index that provides constant returns (and still compounds to the total that was observed).  \\

To better appreciate the different interpretations of these two statistics, let us refer to some examples from our dataset.  In Table \ref{fig_table_sameSN2diffSMC_3} we have two 252 day samples with nearly identical observed SN2s.  Notice the substantial difference in observed SMC, however.  For the TNA sample ending 2/2/2009, the underlying index (Russell 2000) lost about 10\%.  Based on this return the hypothetical best performance of TNA for that same period would be the $R_{MAX}$ of -28.6\%.  The observed return on TNA for that sample period was extremely close, resulting in a very low observed SMC of 2.27\%.  Contrast this with the sample period ending 12/31/2008 that observed a nearly identical SN2 as the TNA sample (0.4220 compared to 0.4148).  Here the index returned 20\% which implies the hypothetical best performance of TNA was 72.01\%.  TNA performed much worse than this maximum (51.84\%) resulting in an observed SMC of 13.28\%.  \\

\begin{table}[ht]
\centering
\scalebox{1}{
\begin{tabular}{rrrr|rrrrr}
 LETF & Leverage & Index & EndDate & SN2 & SMC & $R_{Index}$ & $R_{LETF}$ & $R_{MAX}$ \\ 
  \hline
\hline
TNA & +3 & RTY & 2009-02-02 & 0.4148 & 0.0227 & -0.0990 & -0.2860 & -0.2698 \\ 
   \hline
TNA & +3 & RTY & 2008-12-31 & 0.4220 & 0.1328 & 0.2000 & 0.5184 & 0.7201 \\ 
  \end{tabular}
}
\caption{252 Day Samples with nearly identical observed SN2} 
\label{fig_table_sameSN2diffSMC_3}
\end{table}

Table \ref{fig_table_sameSN2diffSMC_2} provides another pair of samples with nearly identical observed SN2s, but this time we consider 252 day sample periods.  For the EDC sample ending 9/21/2012, the return was about 40\% when the maximum return implied by the underlying index (MSCI Emerging Markets) performance was 53.64\%, resulting in an observed SMC of 9.66\%.  That value is much lower than the observed SMC of 73.80\% that we have for the sample period ending 5/6/2010.  For that sample, the EDC return of 45.82\% was substantially lower than the $R_{MAX}$ of 153.45\%.  \\

\begin{table}[ht]
\centering
\scalebox{1}{
\begin{tabular}{rrrr|rrrrr}
 LETF & Leverage & Index & EndDate & SN2 & SMC & $R_{Index}$ & $R_{LETF}$ & $R_{MAX}$ \\ 
  \hline
\hline
EDC & +3 & MXEF & 2012-09-21 & 0.8212 & 0.0966 & 0.1540 & 0.4010 & 0.5364 \\ 
   \hline
EDC & +3 & MXEF & 2010-05-06 & 0.8250 & 0.7380 & 0.3639 & 0.4582 & 1.5345 \\ 
  \end{tabular}
}
\caption{252 Day Samples with nearly identical observed SN2} 
\label{fig_table_sameSN2diffSMC_2}
\end{table}


For both pairs of sample periods, the observed SN2, or 2-norm of the centered, log returns was nearly identical.  Using this statistic we would interpret the sample periods for each pairing to have exhibited the same realized volatility of returns.  However, within each pair we see substantial differences in shortfall from maximum convexity.  In other words, the sample period with greater observed SMC has a greater degree of underperformance with respect to the sample (event) that observes the underlying index providing a constant daily return (and still compound to the same return for the period).  \\

SMC is valued as a rate of return so the significance of the different observations (within the pairs) can be easily understood by an investor.  If only considering SN2, not only do the observed values lack any kind of real-world interpretation, but there is virtually no measurable difference in realized volatility within the paired sample periods.  We claim that there is a difference in the realized volatilities, and that SMC intuitively quantifies it.  \\

\subsection{Statistical Motivation for SMC}

Now that we have made a qualitative argument for measuring the realized volatility of LETF returns, let us take a hard look at the data.  We survey the sampling distributions of SMC and SN2 across all 20 LETFs in our data set, and using sample statistics provide a quantitative basis for the advantages of our new statistic. \\


Figure \ref{fig:fig_vishist_smcvsSN2} provides box plots of the sample distributions for SMC and SN2 for every long and short LETF using 252 day sample periods.  Looking at the plots for LETFs with underlyer GDM (NYSE ARCA Gold Miners Index) and those with underlyer INDU (Dow Jones Industrials), we see that there is considerable variety within the data set.  We observe that SMC sampling distributions exhibit greater skewness compared to the respective SN2 distribution, as indicated by the length of the start of the right whisker to the end of the outlier points relative to the length of the rest of the box plot.  The median, and thus lower half of each SMC box plot is to the left of the median of the respective SN2 box plot, and on an absolute basis.  Our interpretation of these differences in box plots is that SN2 over-values the intensity of low volatility sample periods and under-values the intensity of high volatility sample periods compared to SMC.  \\

In the previous section we discussed the inequality of SMC and SN2 when calculated on the same series of returns with equal but opposite leverage multiples, and we do see evidence of those relationships in the box plots.  The distributions for SN2 across sides for the same underlying index are very similar, with the only visible difference being the right tail of the plots for index MVRSX.  The differences in plots across sides are much greater with SMC.  The short side box plots for SMC sampling distributions generally have greater ranges, longer right tails, and greater median values compared to the plots for the long side LETF on the same index.  According to SMC we have evidence that LETF pairs on the same index have different risk profiles.  For example, SPXL appears to be more risky than SPXS.  This is noteable because the box plots presented use real LETF returns, which includes a separate observation series for each long side and short side tracking error, and for LETFs with underlying index MVRSX, MXEA, and MXEF those errors were shown to be rather significant.  But our claims in the previous sections assumed we observe LETF returns without noise ($R_{LETF} = \beta R_{Index}$), so we interpret this to mean that the observed values for both volatility statistics are mostly unaffected by tracking error.  This is a point of emphasis in our next section when we use our proposed simulation method with LETF returns.  \\

Range, variance, skewness, and kurtosis of all the SN2 and SMC sample distributions for all LETF tickers are presented in Tables \ref{fig_vishist_table_long} and \ref{fig_vishist_table_short}.  The measures of skewness and kurtosis used are the standardized third and fourth central moments, respectively.  The greater value for each pair of statistics is highlighted in red and we interpret the greater values as evidence that the given statistic provides a more granular level of detail about the realized volatility of the given LETF ticker.  When considering range and variance this interpretation should be obvious--a statistic with a wider range of outcomes or greater variability of outcomes (with respect to the mean) provides more information about the underlying random process.  A greater absolute value of skewness implies longer, fatter tails on one side of the distribution.  We say absolute value of skewness to mean taking the absolute value of the final value for skewness as to allow for negative or positive skew in our comparisons, and this is not to be confused with calculating the third absolute moment.  A greater value for kurtosis implies greater probability mass in the center and tails of the distribution.  In both instances, a greater value implies more mass in the tails and thus greater variability of outcomes.  \\

Across all long and short side LETF tickers the skewness and kurtosis values for SMC are greater than the SN2 values, except for skewness of the pair FINU/FINZ (underlying index Dow Jones US Financials), where both tickers had a negatively skewed sample distribution for SN2.  The variance and range values for SMC are greater than SN2 for most short side tickers, and the results are mixed for long side tickers.  Based on these statistics, our opinion is that SMC provides greater detail about the intensity of realized volatility than SN2 when observing right tail events, and overall SMC is a more informative statistic to use for short side LETF tickers.  \\

%
%


We conclude this section with a brief application example where we rank (separately) the the set of long side and short side LETF tickers by their sample mean SN2 and SMC using 252 day sample periods.  The rankings are displayed in Figure \ref{fig:do_visplot_rank} with rank \#1 having the greatest sample mean for the given statistic, and rank \#10 having the least.  The rank orders that do not agree show two asterisk characters on the tickers at that order number.  \\

There are two sets of two rank orders that do not agree in both the long side and the short side.  In the group of long side LETF tickers, ranks \#2 and \#3, and then ranks \#7 and \#8, do not agree.  On the short side, ranks \#3 and \#4, and also ranks \#7 and \#8, do not agree.  Similar to our observation about the box plots for SN2 displaying minimal differences between long and short side LETF tickers on the same underlying index, here we note that the rankings by SN2 do not change between sides.  The SMC rankings have EDC, RUSL, TNA at \#2, \#3, \#4 on the long side, but on the short side we have RUSS, TZA, EDZ.  The \#7 and \#8 rank orders are the LETFs with the same underlying indexes across long and short sides when using SMC.  \\

Although the sample means for both statistics are plotted on the same axis, we remind the reader that the interpretations of the values are not the same.  short side LETF ticker DUST, for example, has a mean SN2 of about 1.05 and a mean SMC of about 1.3.  SN2 is the 2-norm of the centered, log returns and the mean value of 1.05 offers a general indication of the variability of returns observed in the 252 day sample periods.  SMC is the shortfall of the observed LETF return from a maximum return, so the mean value of 1.3 tells us that the LETF ticker DUST underperforms the maximum return achievable given the index return by 130\% on average over 252 day sample periods.  We feel that the added intuition that comes with measuring realized volatility with SMC makes those rankings more easily understood and more useful when designing trading strategies for these instruments.  \\

\begin{knitrout}
\definecolor{shadecolor}{rgb}{0.969, 0.969, 0.969}\color{fgcolor}\begin{figure}

{\centering \includegraphics[width=\maxwidth]{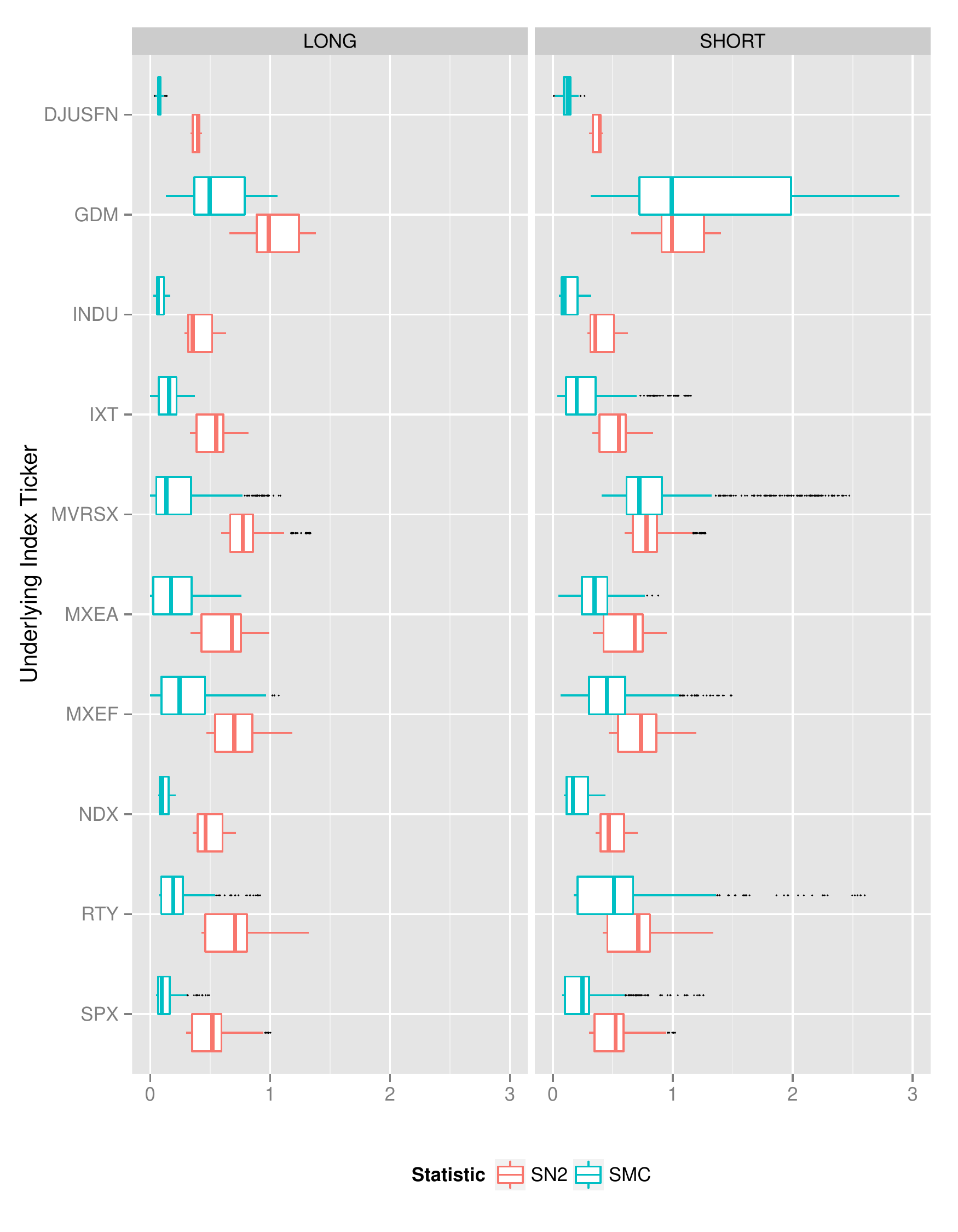} 

}

\caption[SN2 vs]{SN2 vs. SMC for all LETF Ticker Pairs over 252 Day Periods}\label{fig:fig_vishist_smcvsSN2}
\end{figure}

\end{knitrout}
\clearpage

\begin{table}[H]
\centering
{\small
\scalebox{0.8}{
\begin{tabular}{llr|rr|rr|rr|rr}
  \hline
LETF & Index & Obs & Range(SN2) & Range(SMC) & Var(SN2) & Var(SMC) & Skew(SN2) & Skew(SMC) & Kurt(SN2) & Kurt(SMC) \\ 
  \hline
FINU & DJUSFN & 361 & 0.0978 & \textbf{\textcolor{red}{0.1013}} & \textbf{\textcolor{red}{0.0009}} & 0.0002 & \textbf{\textcolor{red}{-0.4852}} & 0.4622 & 1.6312 & \textbf{\textcolor{red}{4.0733}} \\ 
  NUGT & GDM & 762 & 0.7218 & \textbf{\textcolor{red}{0.9325}} & 0.0423 & \textbf{\textcolor{red}{0.0632}} & 0.2626 & \textbf{\textcolor{red}{0.4749}} & 1.7779 & \textbf{\textcolor{red}{1.9898}} \\ 
  UDOW & INDU & 970 & \textbf{\textcolor{red}{0.3475}} & 0.1452 & \textbf{\textcolor{red}{0.0152}} & 0.0018 & 0.6913 & \textbf{\textcolor{red}{0.8978}} & 1.8986 & \textbf{\textcolor{red}{2.1961}} \\ 
  TECL & IXT & 1259 & \textbf{\textcolor{red}{0.4895}} & 0.3731 & \textbf{\textcolor{red}{0.0186}} & 0.0080 & 0.0621 & \textbf{\textcolor{red}{0.2167}} & 1.7703 & \textbf{\textcolor{red}{2.0956}} \\ 
  RUSL & MVRSX & 646 & 0.7468 & \textbf{\textcolor{red}{1.0860}} & 0.0426 & \textbf{\textcolor{red}{0.0669}} & 1.2437 & \textbf{\textcolor{red}{1.5168}} & 3.6069 & \textbf{\textcolor{red}{4.4666}} \\ 
  DZK & MXEA & 1259 & 0.6592 & \textbf{\textcolor{red}{0.7595}} & \textbf{\textcolor{red}{0.0370}} & 0.0339 & 0.0663 & \textbf{\textcolor{red}{0.5625}} & 1.6694 & \textbf{\textcolor{red}{2.0975}} \\ 
  EDC & MXEF & 1259 & 0.7150 & \textbf{\textcolor{red}{1.0721}} & 0.0338 & \textbf{\textcolor{red}{0.0459}} & 0.4742 & \textbf{\textcolor{red}{0.7596}} & 2.0482 & \textbf{\textcolor{red}{2.9446}} \\ 
  TQQQ & NDX & 970 & \textbf{\textcolor{red}{0.3592}} & 0.1506 & \textbf{\textcolor{red}{0.0145}} & 0.0024 & 0.5799 & \textbf{\textcolor{red}{0.7724}} & 1.8429 & \textbf{\textcolor{red}{2.0746}} \\ 
  TNA & RTY & 1288 & \textbf{\textcolor{red}{0.8963}} & 0.8412 & \textbf{\textcolor{red}{0.0434}} & 0.0202 & 0.4335 & \textbf{\textcolor{red}{1.4309}} & 2.1737 & \textbf{\textcolor{red}{5.8182}} \\ 
  SPXL & SPX & 1288 & \textbf{\textcolor{red}{0.7020}} & 0.4450 & \textbf{\textcolor{red}{0.0239}} & 0.0052 & 0.5260 & \textbf{\textcolor{red}{1.4177}} & 2.4170 & \textbf{\textcolor{red}{4.8865}} \\ 
   \hline
\end{tabular}
}
}
\caption{SN2 and SMC (252 Day Periods) Summary Statistics for Long LETF Tickers} 
\label{fig_vishist_table_long}
\end{table}

\begin{table}[H]
\centering
{\small
\scalebox{0.8}{
\begin{tabular}{llr|rr|rr|rr|rr}
  \hline
LETF & Index & Obs & Range(SN2) & Range(SMC) & Var(SN2) & Var(SMC) & Skew(SN2) & Skew(SMC) & Kurt(SN2) & Kurt(SMC) \\ 
  \hline
FINZ & DJUSFN & 361 & 0.1134 & \textbf{\textcolor{red}{0.2573}} & 0.0014 & \textbf{\textcolor{red}{0.0018}} & \textbf{\textcolor{red}{-0.7110}} & -0.0770 & 1.8092 & \textbf{\textcolor{red}{3.0532}} \\ 
  DUST & GDM & 762 & 0.7476 & \textbf{\textcolor{red}{2.5740}} & 0.0442 & \textbf{\textcolor{red}{0.5799}} & 0.2558 & \textbf{\textcolor{red}{0.7119}} & 1.8178 & \textbf{\textcolor{red}{2.0514}} \\ 
  SDOW & INDU & 970 & \textbf{\textcolor{red}{0.3379}} & 0.2677 & \textbf{\textcolor{red}{0.0144}} & 0.0084 & 0.6652 & \textbf{\textcolor{red}{0.8569}} & 1.8727 & \textbf{\textcolor{red}{2.1631}} \\ 
  TECS & IXT & 1259 & 0.5072 & \textbf{\textcolor{red}{1.1146}} & 0.0183 & \textbf{\textcolor{red}{0.0399}} & 0.0818 & \textbf{\textcolor{red}{1.8960}} & 1.8787 & \textbf{\textcolor{red}{7.5079}} \\ 
  RUSS & MVRSX & 646 & 0.6740 & \textbf{\textcolor{red}{2.0666}} & 0.0369 & \textbf{\textcolor{red}{0.2485}} & 1.0722 & \textbf{\textcolor{red}{1.7026}} & 3.1842 & \textbf{\textcolor{red}{4.6711}} \\ 
  DPK & MXEA & 1259 & 0.6181 & \textbf{\textcolor{red}{0.8312}} & \textbf{\textcolor{red}{0.0350}} & 0.0237 & -0.0366 & \textbf{\textcolor{red}{0.5442}} & 1.6074 & \textbf{\textcolor{red}{2.6469}} \\ 
  EDZ & MXEF & 1259 & 0.7309 & \textbf{\textcolor{red}{1.4269}} & 0.0340 & \textbf{\textcolor{red}{0.0577}} & 0.3779 & \textbf{\textcolor{red}{1.0513}} & 2.0376 & \textbf{\textcolor{red}{4.0228}} \\ 
  SQQQ & NDX & 970 & \textbf{\textcolor{red}{0.3544}} & 0.3468 & \textbf{\textcolor{red}{0.0139}} & 0.0136 & 0.5404 & \textbf{\textcolor{red}{0.7388}} & 1.8162 & \textbf{\textcolor{red}{2.0355}} \\ 
  TZA & RTY & 1288 & 0.9188 & \textbf{\textcolor{red}{2.4273}} & 0.0459 & \textbf{\textcolor{red}{0.1466}} & 0.4235 & \textbf{\textcolor{red}{1.6939}} & 2.1804 & \textbf{\textcolor{red}{7.5162}} \\ 
  SPXS & SPX & 1288 & 0.7196 & \textbf{\textcolor{red}{1.1781}} & 0.0239 & \textbf{\textcolor{red}{0.0304}} & 0.5282 & \textbf{\textcolor{red}{2.0606}} & 2.5240 & \textbf{\textcolor{red}{9.9243}} \\ 
   \hline
\end{tabular}
}
}
\caption{SN2 and SMC (252 Day Periods) Summary Statistics for Short LETF Tickers} 
\label{fig_vishist_table_short}
\end{table}

\begin{knitrout}
\definecolor{shadecolor}{rgb}{0.969, 0.969, 0.969}\color{fgcolor}\begin{figure}

{\centering \includegraphics[width=\maxwidth]{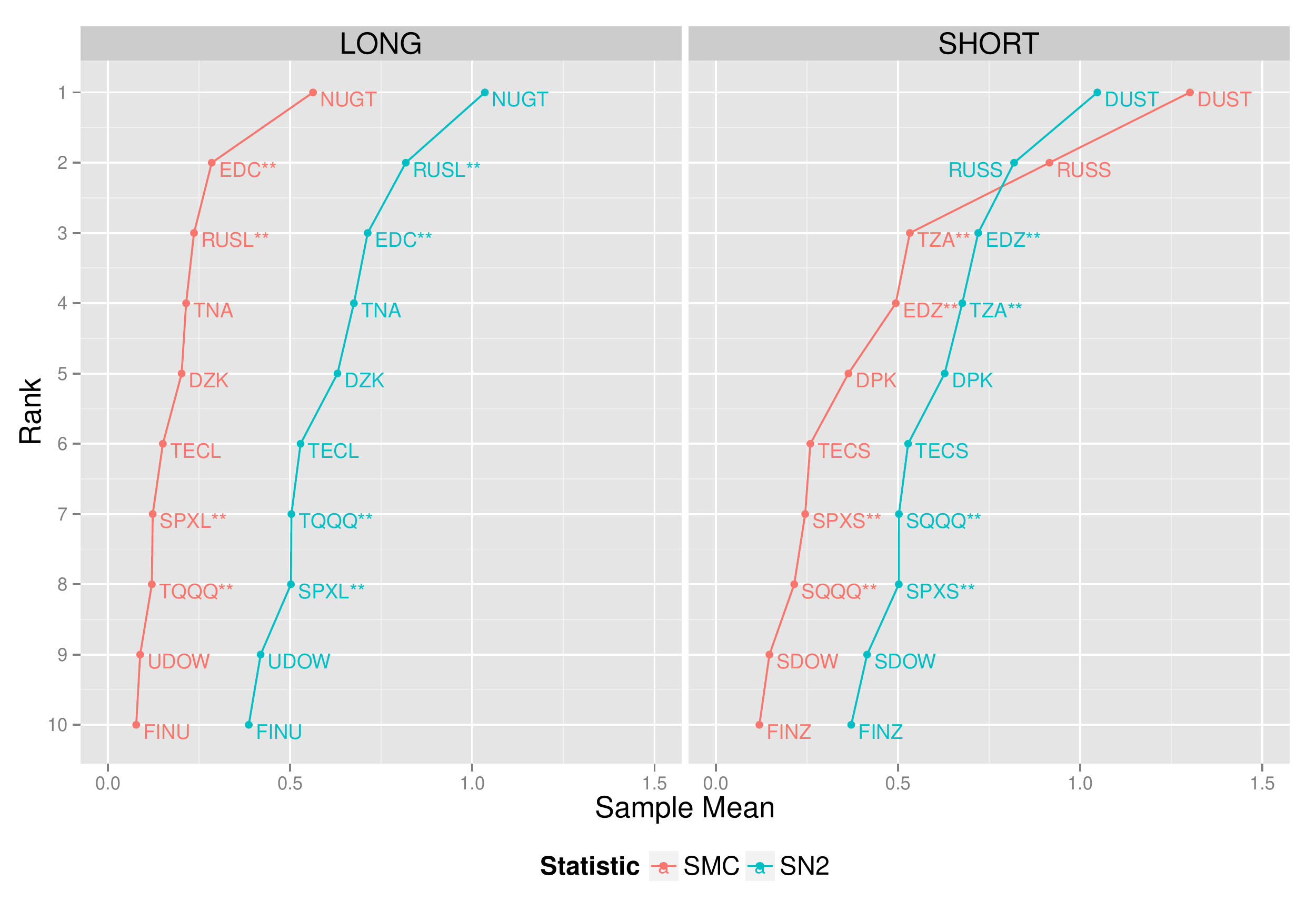} 

}

\caption[Rankings for Long, Short LETF Tickers by SN2 and SMC]{Rankings for Long, Short LETF Tickers by SN2 and SMC}\label{fig:do_visplot_rank}
\end{figure}

\end{knitrout}

%
\clearpage

\bibliographystyle{abbrv}
\bibliography{letf_paper.bib}
\end{document}